\documentclass[reprint,aps,pre,showpacs,superscriptaddress,
amsmath,amssymb,
]{revtex4-1}

\usepackage{graphicx}% Include figure files
\usepackage{dcolumn}% Align table columns on decimal point
\usepackage{bm}% bold math
\usepackage{array}
\usepackage{xcolor}

\begin{document}

\title{The Effects of Communication Burstiness on Consensus Formation and Tipping Points in Social Dynamics}
%\thanks{A footnote to the article title}%

\author{C. Doyle}
\email{doylec3@rpi.edu}
\affiliation{Department of Physics, Applied Physics, and Astronomy \\
Rensselaer Polytechnic Institute, 110 8$^{th}$ Street, Troy, NY 12180--3590, USA}
\affiliation{Network Science and Technology Center \\
Rensselaer Polytechnic Institute, 110 8$^{th}$ Street, Troy, NY 12180--3590, USA}
%\affiliation{Social and Cognitive Networks Academic Research Center}

\author{B.K. Szymanski}
\email{szymab@rpi.edu}
\affiliation{Network Science and Technology Center \\
Rensselaer Polytechnic Institute, 110 8$^{th}$ Street, Troy, NY 12180--3590, USA}
\affiliation{Faculty of Computer Science \& Management\\
Wroclaw University of Science and Technology, 50–-370 Wroclaw, Poland}
%\affiliation{Social and Cognitive Networks Academic Research Center}

%\author{G. Korniss\footnote{Corresponding author. korniss@rpi.edu}}
\author{G. Korniss}
\email{korniss@rpi.edu}
\affiliation{Department of Physics, Applied Physics, and Astronomy \\
Rensselaer Polytechnic Institute, 110 8$^{th}$ Street, Troy, NY 12180--3590, USA}
\affiliation{Network Science and Technology Center \\
Rensselaer Polytechnic Institute, 110 8$^{th}$ Street, Troy, NY 12180--3590, USA}
%\affiliation{Social and Cognitive Networks Academic Research Center}

%\collaboration{Social Cognitive Networking Academic Research Center}
%\noaffiliation

\date{\today}

\begin{abstract}
Current models for opinion dynamics typically utilize a Poisson process for speaker selection, making the waiting time between events exponentially distributed. Human interaction tends to be bursty, though, having higher probabilities of either extremely short waiting times or long periods of silence. To quantify the burstiness effects on the dynamics of social models, we place in competition two groups exhibiting different speakers' waiting-time distributions. These competitions are implemented in the binary Naming Game, and show that the relevant aspect of the waiting-time distribution is the density of the head rather than that of the tail. We show that even with identical mean waiting times, a group with a higher density of short waiting times is favored in competition over the other group. This effect remains in the presence of nodes holding a single opinion that never changes, as the fraction of such committed individuals necessary for achieving consensus decreases dramatically when they have a higher head density than the holders of the competing opinion. Finally, to quantify differences in burstiness, we introduce the expected number of small-time activations and use it to characterize the early-time regime of the system.
\end{abstract}


% Classification Scheme.
%\keywords{Suggested keywords}
%Use showkeys class optizon if keyword
%display desired

\maketitle

\section{Introduction}
Over the years many different models for social dynamics \cite{Castellano_review2009} have been studied on various networks in an attempt to capture different aspects of human behavior \cite{Galam2008, Castellano2003, Steels1995, Baronchelli2006, DallAsta2006, Baronchelli2008,Lu2009}. Some specific models, like the Naming Game (NG) \cite{Steels1995, Baronchelli2006}, have become quite common in the fields of statistical physics and applied mathematics due to their simple rules while keeping some essential features of social dynamics. Unfortunately, the need for simplicity limits the accuracy of these models as they try to balance realism and efficiency. A deeper understanding of the various sources of inaccuracies present is obviously needed, but in the case of social systems gaining this understanding can be quite complex. In many cases, though, alterations of the underlying assumptions and rules within the models can not only help to create a more accurate process, but may allow researchers to draw inferences about the dynamics of social interactions they are attempting to model.

For instance, in the binary Naming Game \cite{Xie2011,Xie2012,Castello2009} (a two-word variant of the Naming Game \cite{Baronchelli2006,Steels1995,DallAsta2006,Baronchelli2008,Lu2009}), the network is made up of many interconnected nodes that each have a list containing one or both of the possible words. A single time step begins with the selection of a speaker followed by a listener. Conventionally, the speaker is chosen randomly, then the listener is chosen again randomly but only from the speaker's neighbors. The speaker then draws a random word within its list and shares that word with the listener. If the listener has that word in its list already, both the speaker and listener delete all other words from their lists. Otherwise, the listener simply adds the speaker's word to its list. The random selection of speakers means that nodes are selected for events via a Poisson process, leading to an exponential distribution of the waiting times they experience between speaking events \cite{Karlin_Taylor1975}. The Poisson communication pattern, however, lacks the richness of realistic communication dynamics \cite{Candia2008,Barabasi2005,Vazquez2006}. In fact, recent works show that human interaction occurs in a far more bursty manner \cite{Goh2008,Karsai2012}; humans tend to speak very frequently for short bursts then go silent for long periods of time, while the exponential distribution leads to fairly regular waiting times (each node will speak on average once every $N$ micro time steps).
A recent study considered the impact of bursty communications on the time to reaching the absorbing state in the voter and in the SI models, where all agents exhibit the same non-Poisson communication characteristics \cite{Artime2016}. In the present work, after introducing the models and methods (Sec.~II), we focus on three scenarios: ({\it i}) Opinion dynamics with competing populations (one with Poisson, the other with bursty communication features) in the binary Naming Game (Sec.~III.A) [and for comparison, in the voter model (Appendix A)]; ({\it ii}) The impact of committed individuals \cite{Xie2011,Xie2012,Zhang2011,Doyle2016,Galehouse2014,Liu2012,Thompson2014,Lu2009,Zhang2012,Mobilia2003,Mobilia2007,Galam2007,Yildiz2011,Marvel2012,Verma2014,Waagen2015} with bursty communication features in the binary NG (Sec.~III.B); ({\it iii}) The impact of committed agents in the base-line scenario where all agents exhibit the same type of bursty communication features (Appendix B).
%%%%%%%%%%%%%%%%%%%%%%%%%%%%%%%%%%%%%%%%%%%%%%%%%%%%%%%%%%%%%%%%%%%%%%%%%%%%%%%%%%%%%%%%%%%%%%%%%%%%%%%%%%%%%%%%%%%%%%%%%%%%%%%%%%%%%%%%%%%%%%%%%%%
To gain some insight into the impact of burstiness, we carried out an analytic approximation for the expected small-time activations of the waiting-time distribution (Sec.~IV).
%%%%%%%%%%%%%%%%%%%%%%%%%%%%%%%%%%%%%%%%%%%%%%%%%%%%%%%%%%%%%%%%%
Further, to broaden the applicability of our work, we demonstrate that our main findings hold for real social networks by studying opinion competition and the impact of committed agents in sparse random graphs (Appendix C).
%%%%%%%%%%%%%%%%%%%%%%%%%%%%%%%%%%%%%%%%%%%%%%%%%%%%%%%%%%%%%%%%%%%%%%%%%%%%%%%%%%%%%%%%%%%%%%%%%%%%%%%%%%%%%%%%%%%%%%%%%%%%%%%%%%%%%%%%%%%%%%%%%%%%

The effects of using more bursty communication patterns have been shown to cause great changes in the properties of models meant to capture the spread of information and diseases, sometimes fascillitating and sometimes slowing spreading depending on the base system and type of network used
\cite{Vazquez2007,Karsai2011,Iribarren2009,Lambiotte2013,Takaguchi2013,VanMieghem2013,Artime2016}. In addition, related studies have shown that increasing the propensity for committed nodes to speak lowers the number of them needed to achieve consensus \cite{Mistry2015}. In social dynamics, however, there is little understanding of how the specifics of the waiting-time distributions may affect the spreading process or what the critical features of the distributions are.
%%%%%%%%%%%%%%%%%%%%%%%%%%%%%%%%%%%%%%%%%%%%%%%%%%%
In contrast to prior works \cite{Vazquez2007,Karsai2011,Iribarren2009,Lambiotte2013,Takaguchi2013,VanMieghem2013,Artime2016}, we investigate this by focusing on studying the effects and impact of agents exhibiting bursty communication delivery {\em competing} with those with Poisson characteristics within the same network. While all individuals have identical speaker-event frequency in the long-time limit, the difference in their burstiness can have profound impact on the opinion competition and consensus formation.

%%%%%%%%%%%%%%%%%%%%%%%%%%%%%%%%%%%%%%%%%%%%%%%%%%%
Using this method to study direct competition between two different inter-event time distributions, we can better see exactly what features of the distributions have the greatest impact on the outcome. Additionally, by combining this competition with the committed agent variant of the naming game we can begin to understand what conditions are most favorable to real world spreading phenomenon. For instance, our results help to inform on what inter-activity time distribution a group of activists should choose to influence a campaign to the largest extent assuming that the rest of the population uses the exponential distribution by default.
\section{Methods}

\subsection{Models}
In order to create competition between nodes following the standard Poisson selection process and those that do not, a set of non-exponential waiting-time distributions can be designed so that each has a mean of one (the same as the exponential distribution generated from the Poisson selection process). Doing so makes the groups identical in frequency of speaking over long times, but different in when they speak. A mean waiting time of one between speaking events also allows for the definition of a single system time step to be such that, on average, there will be $N$ speaking events per unit time. First for the binary NG (Sec.~III.A) [and for comparison, in the voter model (Appendix B)], we perform simulations on complete graphs (fully-connected networks) with the initial condition that half of the nodes have one opinion ($B$) and the standard Poisson speaker selection, while the other half hold another opinion ($A$) and use one of the non-exponential waiting-time distributions. To simulate certain communication patterns as a property specific to individuals, the nodes keep their communication patterns as the system evolves, but their opinions still change in accordance with the binary NG or voter rules, respectively. (I.e., in our models, speaker's inter-event time distribution is a characteristic of a the nodes, not that of the opinion.) Further, to capture some of the effects of sparse social networks, we repeated our analysis of the binary NG on Erd\H{o}s-R\'{e}nyi Random Graphs \cite{ER1960} (Appendix C).

\subsection{Non-Exponential Speakers' Waiting-Time Distributions}
The specific non-exponential distributions chosen for study here can be seen in Table \ref{pdf_dec}. The distributions were chosen largely to reflect the power law nature observed in human communication patterns \cite{Karsai2012,Vazquez2006}, with the Weibull \cite{Johnson1994,White1964} and the uniform distributions used as a controls.
\begin{table*}
	\begin{tabular}{c|c|c|c}
		Name & PDF & Definitions & Restrictions \\ \hline
		Lower cutoff power law & $p(x)=\gamma a^{\gamma} x^{-(\gamma+1)}$ & $a=\dfrac{\gamma-1}{\gamma}$ & $\gamma > 1$, $x > a$ \\ [.5cm]
		Shifted power law & $p(x)=\gamma a^{\gamma} (x+a)^{-(\gamma+1)}$ & $a=\gamma-1$ & $\gamma>1$, $x>0$ \\ [.5cm]
		Weibull & $p(x)=\dfrac{\alpha}{\beta} (\dfrac{x}{\beta})^{\alpha-1} \exp(-(\dfrac{x}{\beta})^\alpha)$ & $\beta=(\Gamma(1+1/\alpha))^{-1}$  & $x>0$ \\ [.5cm]
		Uniform & $p(x)=1/b$ &   & $1-b/2 < x < 1+b/2$, $b<2$ \\ [.5cm]
	\end{tabular}
	\caption{Description of the probability density functions for the different non-exponential speakers' waiting-time distributions. The parameters $\gamma,\alpha, b$ are used to control the burstiness of the distributions.}
	\label{pdf_dec}
\end{table*}

\begin{figure*}
		\centering
		\includegraphics[width=.75 \textwidth]{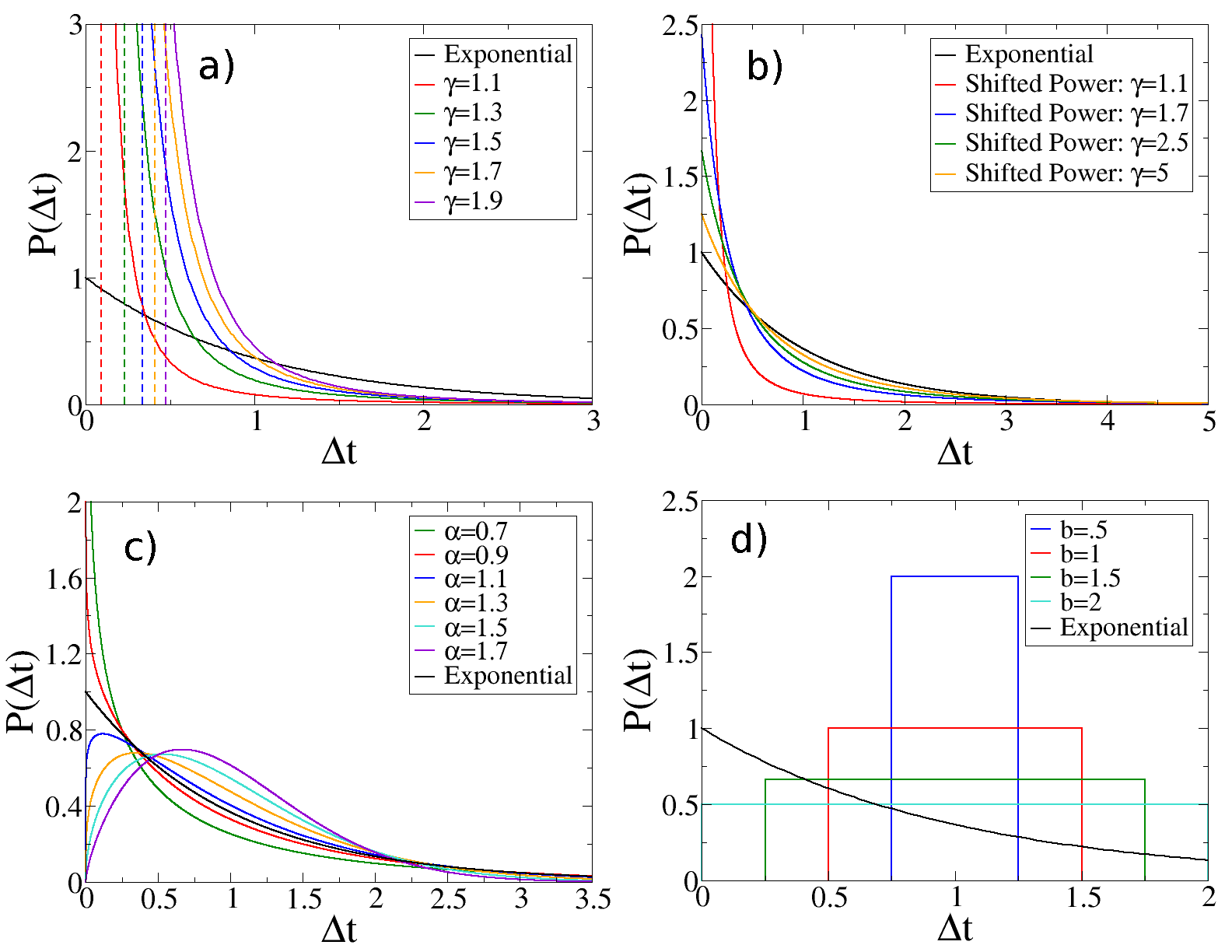}
		\caption{PDFs for the non-exponential speakers' waiting-time distributions with various chosen parameters compared to the exponential one. (a) the power law with lower cutoff, (b) the shifted power law, (c) the Weibull distribution, and (d) the uniform distribution.}
		\label{pdf_plot}
	\end{figure*}

In Fig.~\ref{pdf_plot}, the different probability density functions (PDFs) for each of the distributions can be seen along with the PDF of the exponential distribution. These plots provide the basis for a qualitative understanding of why each distribution is used, as well as providing an intuition for how each distribution behaves as both are needed to explain results going forward. First, Fig.~\ref{pdf_plot}(a) shows the power law with a lower cutoff at $a=(\gamma-1)/{\gamma}$. This distribution was chosen for its propensity towards burstiness, but also the regularity caused by the short time dead period. The cutoff means that there is a minimum time each node must wait between speaking events, and as the system gets burstier (small values of $\gamma$) the cutoff grows. Then, to test the behavior of a system with an unrestricted bursty nature, Fig.~\ref{pdf_plot}(b) shows the behavior of a power law translated to the left by the value $a=\gamma-1$. This system always maintains a higher head density and is thus always burstier than the exponential distribution (though it becomes similar to the system with exponential distribution for large values of $\gamma$). Then, Fig.~\ref{pdf_plot}(c) displays the Weibull function. This function has some behavior derived from both the power law and exponential distributions, and in the special case of $\alpha=1$, it is exactly the exponential function \cite{VanMieghem2013}. For $\alpha<1$ it is always more bursty and for $\alpha>1$ it is always less bursty than the exponential distribution. Lastly, Fig.~\ref{pdf_plot}(d) is a uniform distribution centered around $x=1$ with a range of $b$, a function that is always clearly less bursty than the exponential one.
%%%%%%%%%%%%%%%%%%%%%%%%%%%%%%%%%%%%%%%%%%%%%%%%%%%%%%%%%%%%%%%%%%%%%%%%%%%%%%%%%%%%%%%%%%%%%%%%%%%%

\section{Simulation Results}

\subsection{Opinion Competition in the Binary Naming Game}
\label{ng_compete_sec}
\begin{figure*}
	\centering
	\includegraphics[width=.75\textwidth]{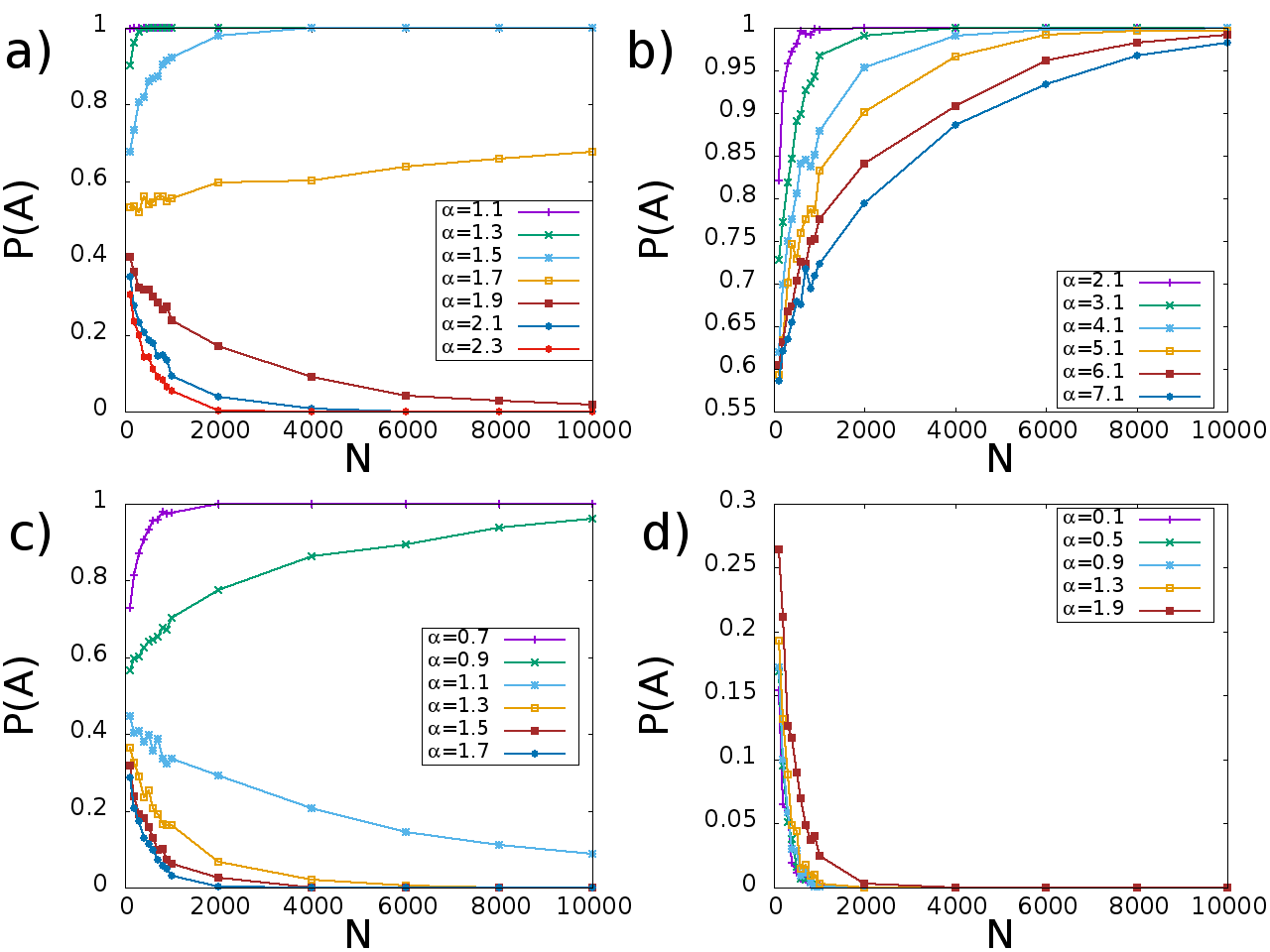}
\caption{The fraction of runs (out of 1000 trials) vs system size that the non-Poisson ($A$-opinion) nodes won the opinion competition against the Poisson ($B$-opinion) nodes in the binary NG on a complete graph. As before, the speakers' waiting-time distribution for the non-Poisson nodes is (a) power law with lower cutoff, (b) shifted power law, (c) Weibull, and (d) uniform distribution.}
	\label{ng_sims}
\end{figure*}

In this section we study simulations of the competition outlined above by running the system to consensus and comparing the fraction of wins for the non-Poisson nodes with initial opinion ($A$) for a given system size and set of control parameters. Here we limit the study to simulations on a complete graph, but we show in Appendix C that the results found do not change when the system is run on a sparse random network instead. As seen in Fig.~\ref{ng_sims}, opinions corresponding to the burstier waiting-time distribution are favored, an effect that becomes more pronounced with increasing system size. In fact, in the case of the power law with lower cutoff and the Weibull distributions, there is a fairly clear transition at large system sizes at $\gamma \approx 1.7$ and $\alpha = 1$ respectively due to those parameters allowing these distributions to have either a higher or lower head density than the exponential distribution. This is particularly interesting for the power law with a lower cutoff. In this case, the transition is the result of the regularity of the forced waiting period between speaking events and the inherent burstiness of the power law head balancing out around $\gamma=1.7$. In both cases, though, the dominance of the burstier distribution becomes more pronounced at large system sizes, causing the side with the higher head density to win with near certainty in large systems. This is further supported by the results of the simulations with the shifted power law and uniform distributions. Since there is no transition of head density in these cases (the shifted power law is always burstier than exponential while the uniform is always less bursty), they are always more and less likely to win, respectively. These results imply that despite the efforts to preserve the symmetry of the system by keeping the mean waiting times the same across all distributions, simply changing the way these waiting times are distributed carries sufficient impact to entirely break the symmetry (in the limit of infinite system size [Fig.~\ref{ng_sims}]).
%%%%%%%%%%%%%%%%%%%%%%%%%%%%%%%%%%%%%%%%%%%%%%%%%%%%%%
This is in contrast to the voter model, studied in Appendix A, where the bias towards the burstier opinion remains constant with increasing system size. In that case, the randomness inherent in the voter model works to mitigate the effect of the early time dominance of the burstier opinion and allows the system to revert to an even competition more easily.

\begin{figure*}
	\centering
	\includegraphics[width=.75\textwidth]{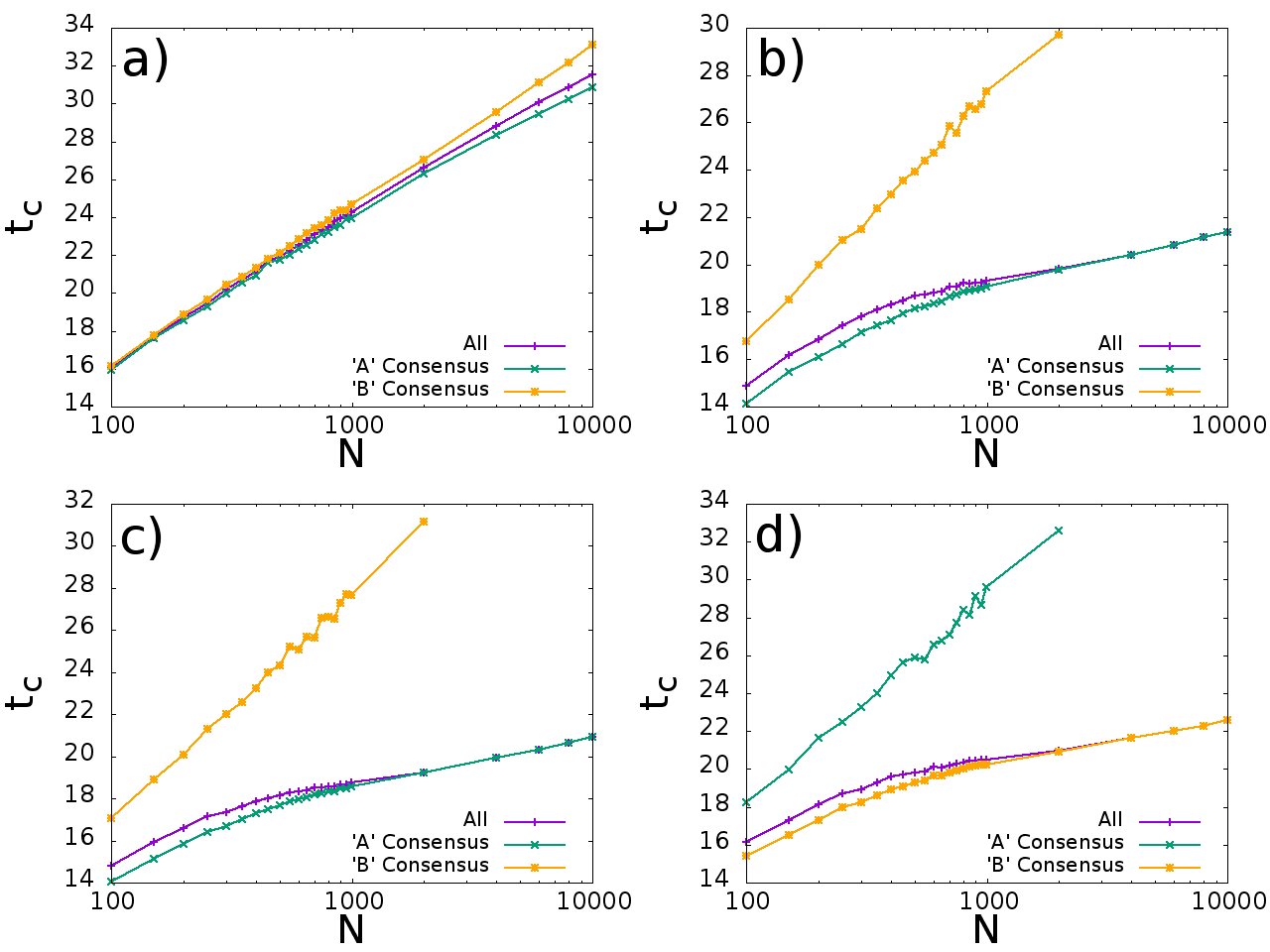}
	\caption{The time to consensus conditioned on each side wining in the binary NG on a complete graph. Initially, half the nodes have non-exponential speakers' waiting-time distribution and hold opinion $A$, while another half follows an exponential distribution and hold opinion $B$. Part (a) displays the results for the power law with a lower cutoff and $\gamma=1.7$ in an attempt to show results for a system with similar burstinesses. Parts (b) and (c) shows the case of a more bursty non-exponential distribution (the shifted power law with $\gamma=2.9$ and the Weibull distribution with $\alpha=0.7$ respectively) while part (d) shows the less bursty case (uniform distribution with $b=1.9$). All simulations were run 10000 times, then averaged based on their final consensus.}
	\label{ng_ct}
\end{figure*}

\begin{figure*}
	\centering
	\includegraphics[width=.75\textwidth]{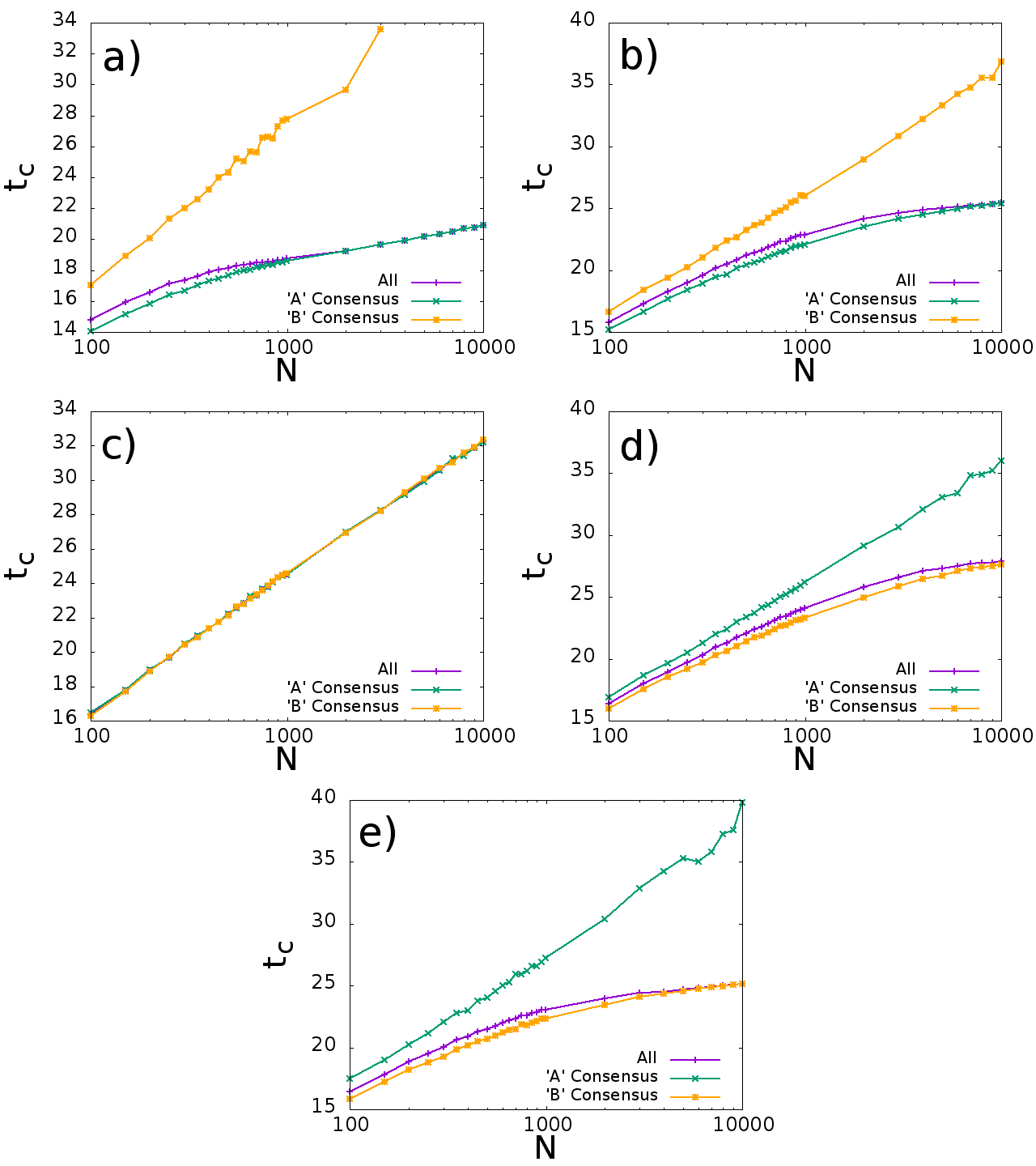}
	\caption{The time to consensus conditioned on each side winning in the binary NG on a complete graph, where half of the nodes are in opinion $A$ initially and follow a Weibull waiting-time distribution and the other half of the nodes initially are in opinion $B$ following an exponential waiting-time distribution. Part (a) shows the results of the conditioned simulations with $\alpha=0.7$ for the Weibull distribution, (b) with $\alpha=0.85$, (c) with $\alpha=1$, (d) with $\alpha=1.15$, and (e) with $\alpha=1.3$.}
	\label{weibull_ct}
\end{figure*}

\begin{figure*}
	\centering
	\includegraphics[width=.75\textwidth]{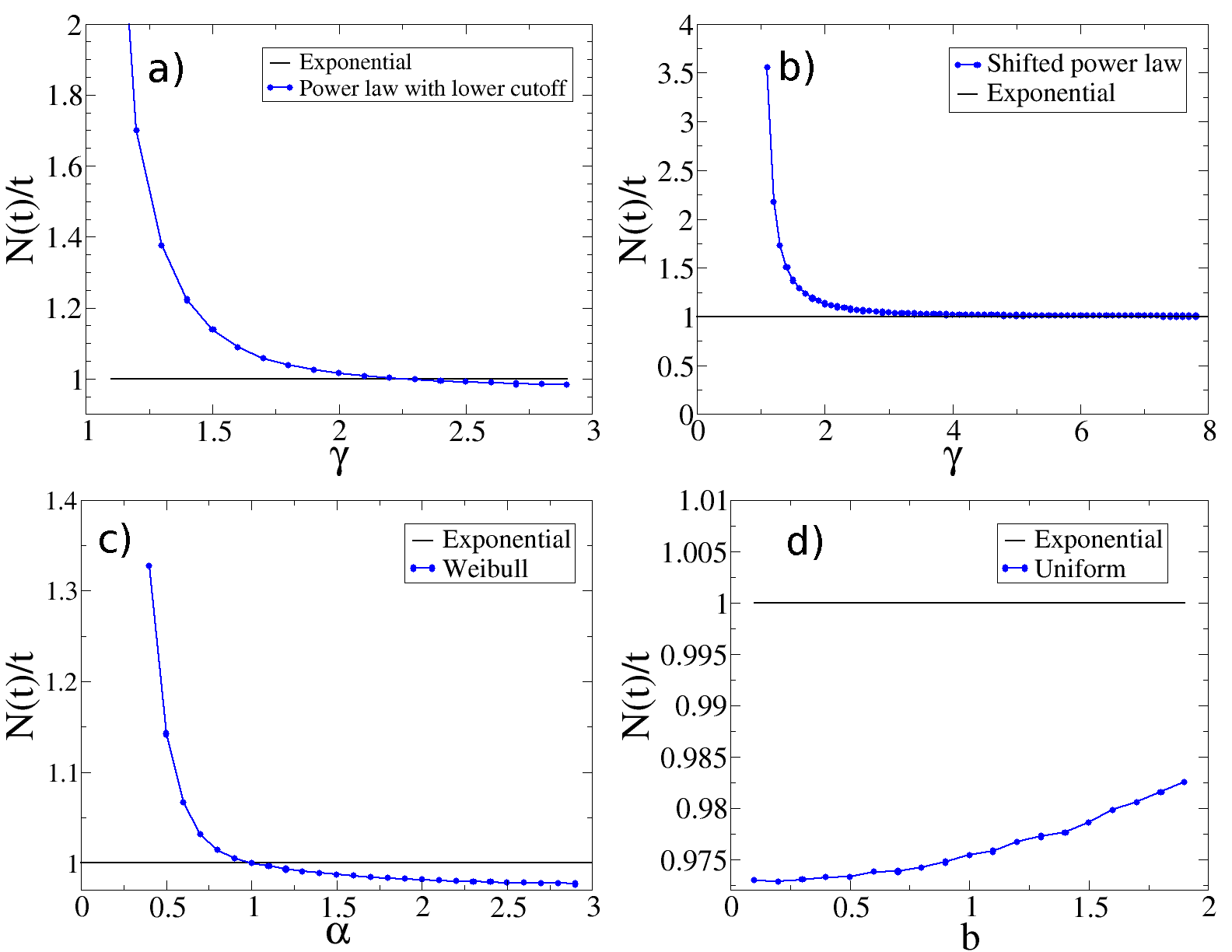}
	\caption{The average number of speaking events in the binary NG for each type of nodes per one system time-step over the course of the entire simulation. Each simulation is averaged over 1000 runs with $N=1000$ on a complete graph. As before, (a) is the power law with lower cutoff, (b) is the shifted power law, (c) is the Weibull, and (d) is the uniform distribution.}
	\label{ng_rate}
\end{figure*}

The question of why these distributions behave this way (and why the head of the distributions matters more in this context than the tail) can be answered by studying the different time regimes of the system. By looking at the average time to consensus for the systems conditioned on which opinion eventually won (as seen in Figs. \ref{ng_ct} and \ref{weibull_ct}), the time scales on which the distributions operate can be seen more clearly. Specifically, the system takes a much longer time to reach consensus for the less bursty opinion than for the more bursty one. This can be explained by dividing the simulations into two time regimes: early time and late time. In the early time regime, the burstier nodes dominate since they are likely to activate (often multiple times) before the less bursty nodes activate at all. They are then likely to go dormant for some extended amount of time, beginning the later time regime where the less bursty nodes become far more active. In most cases, however, the early time dominance of the burstier side switches the opinon for a sufficient number of the less bursty nodes to create a heavy majority for the burstier side before the later time regime is entered. When this happens, the system quickly reaches consensus before many of the nodes even have their opportunity to speak, leading to the heavily unbalanced average activations per time step seen in Fig.~\ref{ng_rate}. This result indicates that a high head density (correlating to a strong initial push of opinions) is critical to achieving consensus, even if the nodes that initially caused the push go silent for long periods afterwards. Even so, occasionally the less bursty opinion still has enough of a presence to push back during the later time regime to allow for the long time victories of the less bursty opinion.
%A more in depth look at the activation rates of nodes in extreme bursty cases and how they react to different time scales is provided in Appendix B.
%%%%%%%%%%%%%%%%%%%%%%%%%%%%%%%%%%%%%%%%%%%%%%%%%%%
We have also found that the consensus time increases {\em logarithmically},  $t_c\sim\ln(N)$, in the asymptotic large system-size limit [Figs.~\ref{ng_ct} and \ref{weibull_ct}]. This logarithmic scaling holds for all cases of competing non-Poisson communication dynamics, and regardless of the outcome of the competition. The rate of the logarithmic increase (i.e., the slope of the lines in the log-normal plots in Figs.~\ref{ng_ct} and \ref{weibull_ct}), however, is sensitive to the details of the non-exponential waiting-time distributions and to the condition of the outcome of the competition. One should also note that the standard binary NG with only Poisson communication dynamics also exhibits logarithmic scaling of the consensus times with the system size \cite{Baronchelli2008,Castello2009,Zhang2011}.

\subsection{Consensus Formation and Tipping Points with Committed Agents in the Binary NG}
In prior works, models with committed agents (or zealots) have often been employed to simulate opinion spread driven by individuals who never change their opinion \cite{Xie2011,Xie2012,Zhang2011,Doyle2016,Galehouse2014,Liu2012,Thompson2014,Lu2009,Zhang2012,Mobilia2003,Mobilia2007,Galam2007,Yildiz2011,Marvel2012,Verma2014,Waagen2015}. The general model with committed agents takes a small minority population in the network and removes their ability to change opinion, but still allows them to communicate and share their opinion with others. In most cases, simulations with committed agents are set up so that a small population of nodes ($p$) within the system are designated committed agents and given a single opinion ($A$). All other nodes in the system follow the rules of the binary NG as usual and are initialized with the other opinion ($B$). Because the committed agents will never change their opinion, the only possible stable state is the consensus on the minority opinion. In the binary NG, however, there exists a critical fraction of committed individuals ($p_c$) (the tipping point) that causes a sharp phase change in the system such that below $p_c$ the system reaches consensus on the order of $T_c\sim \exp(N)$ while above $p_c$ the system reaches fast consensus on the order of $T_c\sim \ln(N)$ \cite{Xie2011}.
%%%%%%%%%%%%%%%%%%%%%%%%%%%%%%%%%%%%%%%%%%%%%%%%%%%%%

\begin{figure*}
	\centering
	\includegraphics[width=.75\textwidth]{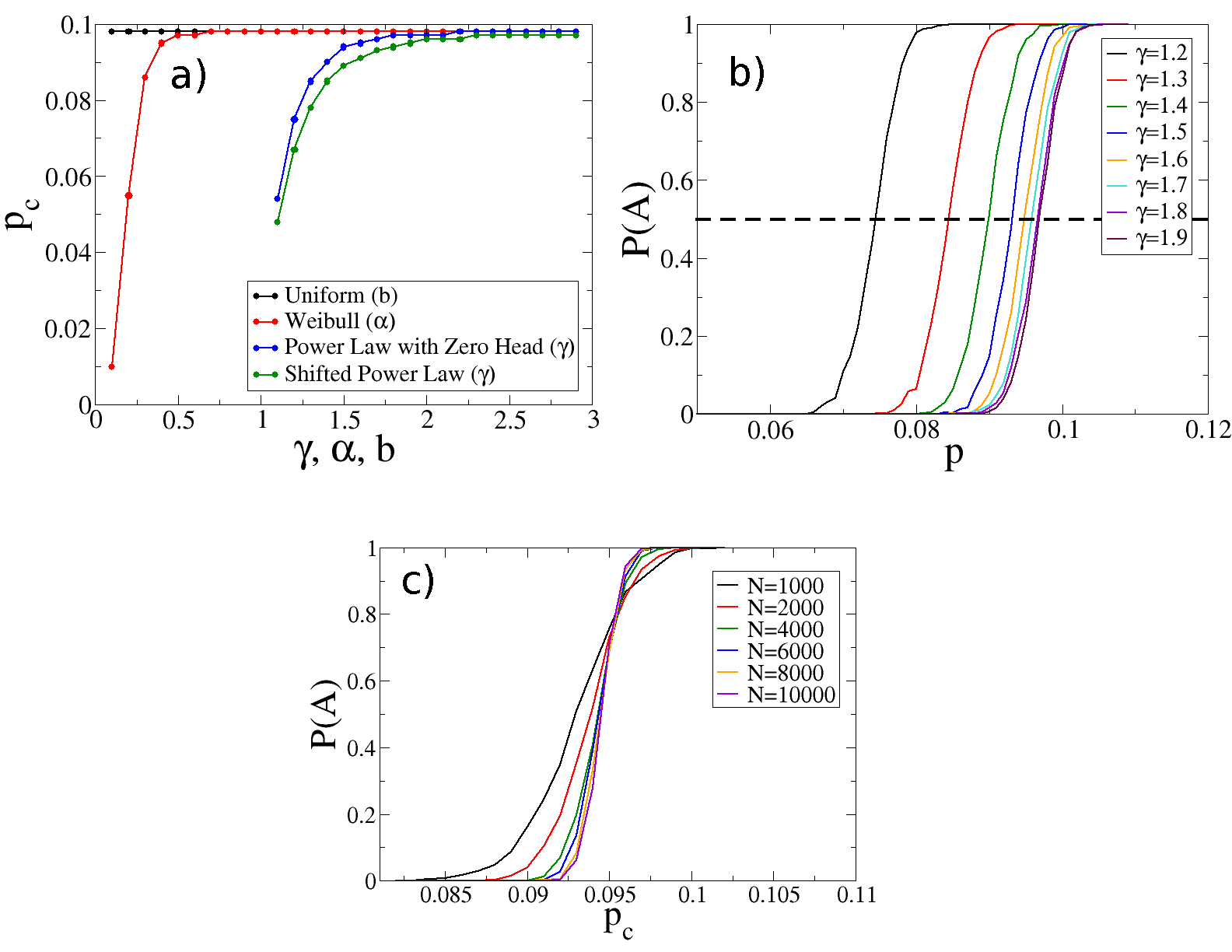}
	\caption{The effects committed agents with non-Poisson speakers' communication patterns in the binary NG on a complete graph.
	(a) The critical fraction of committed agents (tipping point) necessary to create consensus for the minority opinion with respect to the various parameters that control the burstiness of the committed agents' waiting-time distributions, averaged over 1000 runs on systems with $N=1000$.
	It is important to note that the parameters $\gamma$, $\alpha$, and $b$ are specific to the distribution in which they are used and their impact on the burstiness varies from one distribution to another. As such, they should not be compared directly to each other.
	(b) Fraction of runs reaching consensus driven by committed minority by time $t=150$. Committed agents have power law with lower cutoff speakers' waiting time distribution with parameter $\gamma$. The critical fractions $p_c$ [shown in (a)] were defined as the population at which the system reaches consensus (of the initial minority opinion) in over half of the runs. (c) Finite-size effects of the tipping point for  power-law with the lower cutoff and $\gamma=1.5$, indicating no significant shift in the value of $p_c$ as $N \rightarrow \infty$.}
	\label{committed}
\end{figure*}

The value of $p_c$ is somewhat sensitive to small alterations within the system rules or the average node degree \cite{Zhang2012}, and heterogeneous waiting-time distributions are able to lower $p_c$ considerably (as seen in Fig.~\ref{committed}(a)). Here, $p_c$ is defined as the committed population at which half of 1000 simulations reaches consensus before $t=150$, shown in Fig.~\ref{committed}(b). The system size in Fig.~\ref{committed}(a) is 1000, but as Fig.~\ref{committed}(c) shows, there is no shift in $p_c$ at higher values of $N$. In these simulations, only the committed agents are designated as being non-Poisson nodes, while all other nodes followed the standard Poisson selection process. When the committed agents are burstier than the surrounding population, they are able to work far more efficiently and lower the critical fraction of the population considerably. Interestingly, the opposite is not true. When the non-committed nodes are burstier, the critical fraction remains steady at $p_c\approx0.098$. This is due to the same time regime dynamics discussed earlier; the burstier nodes speak frequently in the early time regime and give a heavy advantage to their side. If those burstier nodes are the committed agents, they establish a strong minority presence in the simulation and gain an advantage. If they are not the committed nodes, however, no advantage is gained because the committed nodes cannot change their opinion. Instead, the system enters the long time regime where the identical mean waiting times take over and the system reverts to the value of $p_c$ that occurs in a simulation with all speakers being Poisson selected.

To gain further insight in the impact of bursty communication patterns on the tipping point $p_c$, we also investigated the base-line scenario where
all individuals in the system exhibit the same type of non-exponential waiting-time distribution. The results are shown in Appendix B.

\section{Approximation of the Expected Small-Time Activations}
Throughout the preceding section we demonstrate via direct simulation that competition between two opinions spread by groups with different levels of burstiness will favor the opinion of the group with the higher burstiness. Further, we demonstrate that the relevant quantity of the waiting-time distribution is the head density rather than the tail due to the importance of dominating the initial stages of the simulation. An analytic description of this phenomenon proves difficult, however, because direct comparisons of the head density via the CDF fail to accurately describe the dynamics of this system. These simple comparisons do not sufficiently account for the probability that a bursty node can activate multiple times before a less bursty node activates once, and thus greatly underestimate the effect that burstiness can have on a system. To remedy this, we propose using the \emph{expected small-time activations} ($D$) to characterize the burstiness of a given node. The expected small-time activations is an approximation of how many times a node following a given waiting-time distribution is expected to activate before the mean activation time is reached. This value allows for a direct comparison of the influence that different distributions have over the early time period of a simulation by sampling the head of the distribution multiple times within different ranges to account for a node activating multiple times within this range. Using the notation from before where $p(x)$ is the PDF of the waiting-time distribution, and $P(t)=\int_{0}^{t}p(x)dx$ is the CDF of the same function, we can begin to calculate the expected small-time activations $D$. First, we can give the probability that a node will activate exactly $m$ times before $t$ as
\begin{equation}
P_m=\int_{0}^{t} p(x)P_{m-1}(t-x)dx
\end{equation}
with the special case of no activations before $t$ being $P_0(t)=1-P(t).$ From here, we can begin to approximate $D$ by summing the probabilities that a node will speak $m$ times before $t$ multiplied by $m$. This is continued for all values of $m$ up to a maximum value considered ($n$) after which it is assumed that if a node has not activated $n$ or less times then it must activate exactly $n+1$ times. Thus, we say that the order of the approximation is $n$ and an approximation of order $n$ will consider a maximum number of activations $n+1$. The definition of $D$ of order $n$ is given by
\begin{equation}
	D_n(t)=(n+1)\Big(P(t)-\sum_{m=1}^{n} P_m(t)\Big)+\sum_{m=1}^{n} mP_m(t) \;.
	\label{approx_eq_gen}
\end{equation}
Of course, as $n$ goes to infinity this approximation becomes an exact description of the expected number of activations before $t$. The probability of having $n>3$ activations, however, drops very quickly with $n$, so the expected value of $D$ for a distribution comparable to the exponential (where $D=1$) can be reasonable well-approximated by $D_2$. Hence, for simplicity we consider only this case in this paper, and by following the procedure outlined above the approximate values can be calculated via Eq. (\ref{approx_eq2})
\begin{equation}
	D_2(t)=3P(t)-2P_1(t)-P_2(t)  \;.
	\label{approx_eq2}
\end{equation}

For the exponential and uniform distributions, $D$ can be solved exactly up to higher orders. In fact, the specific values of $P_n$ are well known for the exponential distribution as
\begin{equation}
P^\textrm{exp}_n(t)= \frac{t^n}{n!} e^{-t} \;.
\end{equation}
Similarly, for the uniform distribution the values of $P_1$ and $P_2$ (and beyond) can be obtained analytically in a closed form,
\begin{align}
	\label{uni_p1}
P^\textrm{uni}_1(t)=&~\Theta(t-(1-b/2))\bigg(\frac{t-1+b/2}{b}\bigg)\\
\nonumber&-\Theta(t-2(1-b/2))\bigg(\frac{(t-2+b)^2}{2b^2}\bigg)   \;,
\end{align}
\begin{align}
	\label{uni_p2}
P^\textrm{uni}_2(t)=&~\Theta(t-2(1-b/2))\bigg(\frac{(t-2+b)^2}{2b^2}\bigg)\\
\nonumber& -\Theta(t-3(1-b/2))\bigg(\frac{27b^3+54b^2t}{48b^3}\\
\nonumber& +\frac{36bt^2+8t^3-162b^2-216bt-72t^2}{48b^3}\\
\nonumber& +\frac{324b+216t-216}{48b^3}\bigg) \;,
\end{align}
where $\Theta(t)$ represents the Heaviside step function. It should be noted that Eqs.~(\ref{uni_p1}) and (\ref{uni_p2}) are only valid for $t\le1+b/2$ (including the range of our interest $t\leq1$).

\begin{figure*}
	\centering
	\includegraphics[width=.75\textwidth]{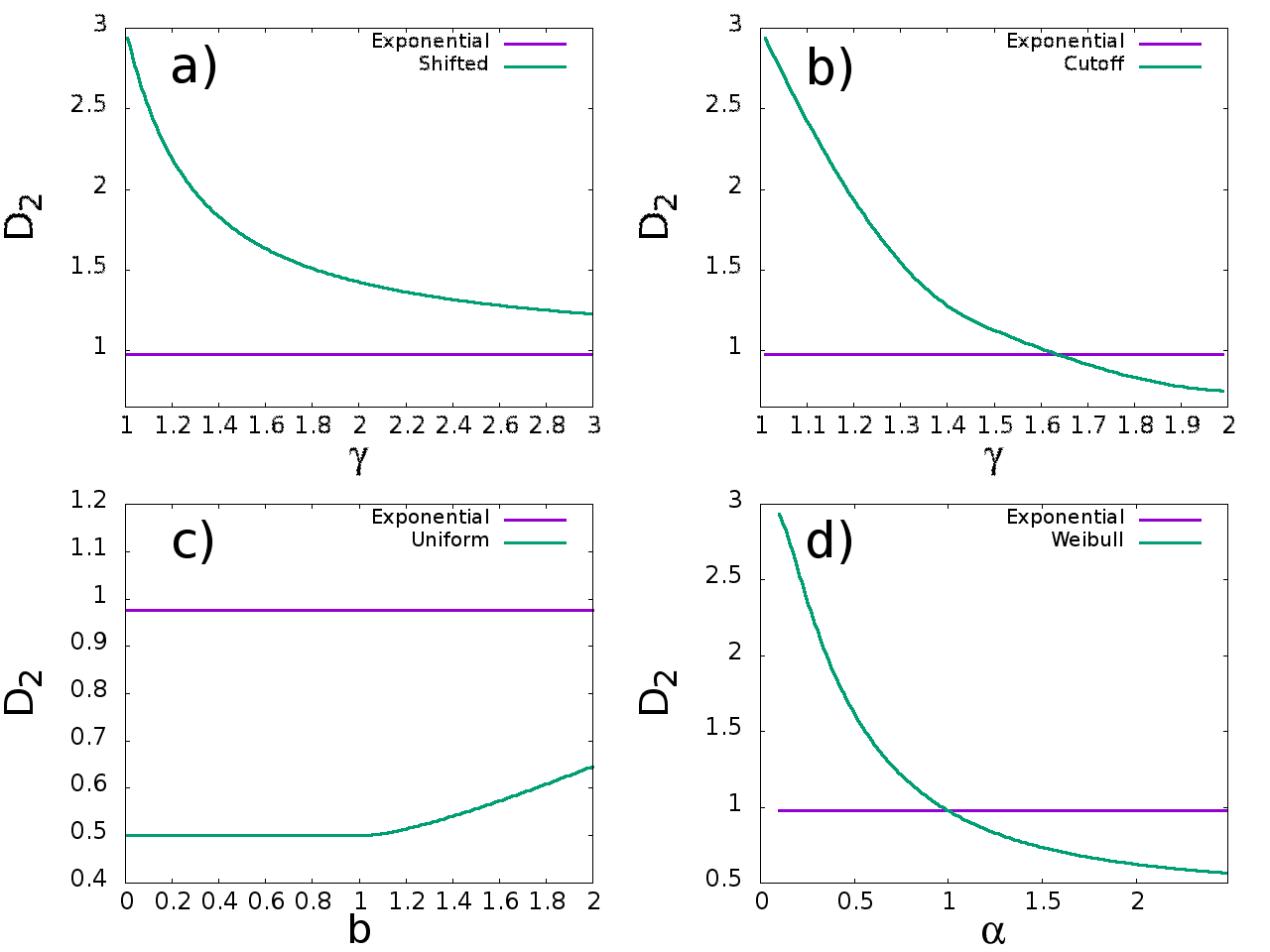}
	\caption{Comparison of the second order approximation of the small-time activation densities for each of the non-exponential distributions vs the exponential. (a) shows the shifted power law, (b) shows the power law with lower cutoff, (c) shows the uniform distribution, and (d) shows the Weibull distribution.}
	\label{approx_figs}
\end{figure*}

Unfortunately, for the other waiting-time distribution functions, the complexity of the integrals limits to which order the approximation can be taken analytically. For instance, both $P_1$ and $P_2$ for the Weibull distribution must be computed numerically, while the values of $P_2$ for both of the power laws also require numeric integration. The first order (using just $P_1$) approximation for each of the power laws can be computed analytically as
\begin{align}
	P^\textrm{shifted}_1=&-\gamma \bigg(\frac{a}{2a+t}\bigg)^{2\gamma} \Bigg(B\bigg(\frac{a}{2a+t};-\gamma,1-\gamma\bigg)\\
	\nonumber&-B\bigg(\frac{a+t}{2a+t};-\gamma,1-\gamma\bigg)\Bigg)
\end{align}
and
\begin{align}
	P^\textrm{cutoff}_1=&~\Theta(t-a)\Big(1-(a/t)^\gamma\Big)\\
	\nonumber&-\Theta(t-2a)\Bigg[1-\bigg(\frac{a}{a-t}\bigg)^\gamma\\
	\nonumber&-\gamma\bigg(\frac{a}{t}\bigg)^{2\gamma}\Bigg(B\bigg(\frac{t-a}{t};-\gamma,1-\gamma\bigg)\\
	\nonumber&-B\bigg(\frac{a}{t};-\gamma,1-\gamma\bigg)\Bigg)\Bigg] \;,
\end{align}
where $B(x;p,q)$ denotes the incomplete beta function $B(x;p,q)=\int_0^x t^{p-1}(1-t)^{q-1}dt$ (see \cite[Eq.~(8.17.1)]{DLMF}). The remaining cases of $P_2$ for the power law distributions and both $P_1$ and $P_2$ for the Weibull distribution require numeric integration as mentioned above. Using these formulas to find approximate values for the expected small-time activations via Eq. (\ref{approx_eq2}) (and using $t=1$) yields the the results in Fig. \ref{approx_figs}, an accurate representation of the approximate burstiness of each distribution with respect to its controlling parameter. In the most simple cases of the shifted power law and the uniform distribution, this means accurately displaying that they are always more or less bursty (respectively) than the exponential. Additionally, the approximation is able to easily predict the point of equal burstiness in the trivial case of the Weibull distribution, since the Weibull becomes exactly the exponential for the case of $\alpha=1$. Finally, and most importantly, the approximation predicts the transition point of the power law with lower cutoff to be $\gamma\approx1.64$, a value in close agreement with the prior simulated results. This agreement with a value produced using only information from the head of the waiting-time distributions further strengthens the assertion that the dominant region of the distribution with relation to the outcome of social simulations is the head density rather than the tail, as contributions from any other regions must be small and make up at most the $<5\%$ difference between the values.
%%%%%%%%%%%%%%%%%%%%%%%%%%%%%%%%%%%%%%%%%%%%%%%%%%%%%%%%%%%%%%%%%%%%%%%%%%%%%%%%%%%%%%%%%%%%%%%%%%%%%%%%%%%%%%%%

\section{Conclusion}

Attempts to bring more human communication patterns to social dynamics models are often difficult, but understanding the effects of different changes helps to bring the abstract models closer to reality. Using a Poisson process to select speakers in pairwise interaction models is popular and extremely attractive for its simplicity, yet it is quite different from how people behave. Not only do people tend to have a much burstier communication patterns, but they tend to act individually and thus heterogeneously. Implementing such dynamics into common pairwise interaction models shows a powerful advantage of the community with burstier communication pattern. In the binary NG, this effect is stronger as the system size increases since the symmetry of the system is broken. This allows even a very small difference in the waiting-time distribution to have a large effect on the outcome for a sufficiently large system. In the voter model, however, this scaling effect is lost, but the overall bias towards the burstier community remains (see Appendix A).

When committed agents are introduced to the models, prior work indicates that the main factors impacting the value of the tipping point were the number (or fraction) of committed agents \cite{Xie2011,Xie2012,Zhang2011}, the level of commitment of those agents \cite{Doyle2016,Galehouse2014}, their eagerness to leave an intermediate opinion state [34], the average node degree [35], and their rate of activation relative to other nodes in the system \cite{Mistry2015}. The results presented here indicate another factor: the waiting-time distribution of the committed agents relative to that of the surrounding nodes, as the general bias towards a more bursty community remains with the presence of committed agents. In fact, the waiting-time distribution effect is particularly interesting in situations where it is desirable to minimize the size of the committed fraction because the heterogeneous waiting-time distributions can have only a positive impact on the efficiency of the committed agents. If the committed agents are less bursty, the system simply enters the long time regime and reverts to the critical fraction for a system of homogeneous nodes all with exponentially distributed waiting times. This effect is important in the study of facilitating  the growth of a single opinion in a society, as it implies a new strategy consisting of multiple strong pushes for the new opinion even if they are separated by long periods of inactivity. Such a pattern can heavily decrease the cost of spreading an opinion throughout a society (assuming cost to be proportional to the number of committed agents rather than to the number of messages sent, otherwise the results indicate that a high investment upfront is desirable) by increasing the efficiency of any concerted effort by activists to aid the spread spread. Unfortunately, analytic study of this process is somewhat limited as the non-Markovian nature of the various selection processes inhibits such endeavors. The phenomenon can, however, be accurately approximated by calculating the expected number of activations before the mean waiting time. This allows for a relatively simple method for comparing the burstiness of different distributions in terms of their impact on the early time period of a simulation. Using this information from only the head of the nodes' waiting-time distributions, accurate predictions on the outcomes of simulations can be obtained as it is clear that systems that are otherwise entirely symmetric can be heavily biased towards one opinion or another via changing the waiting time distribution of a portion of the nodes.

%Additionally, future work is needed to complete analysis of the effects of homogeneous non-exponential waiting-time distributions in the binary NG.

\section*{Acknowledgements}
This research was supported in part by
the Army Research Laboratory under Cooperative Agreement Number W911NF-09-2-0053 (the ARL Network Science CTA), by the Office of Naval Research (ONR) Grant No. N00014-15-1-2640, and by the National Science Centre research project DEC-2013/09/B/ST6/02317. The views and conclusions contained in this document are those of the authors and should not be interpreted as representing the official policies either expressed or implied of the Army Research Laboratory or the U.S. Government.

\newpage

\appendix

\section{Opinion Competition in the Voter Model}
	\label{voter_compete_sec}
\begin{figure*}

		\centering
		\includegraphics[width=.75\textwidth]{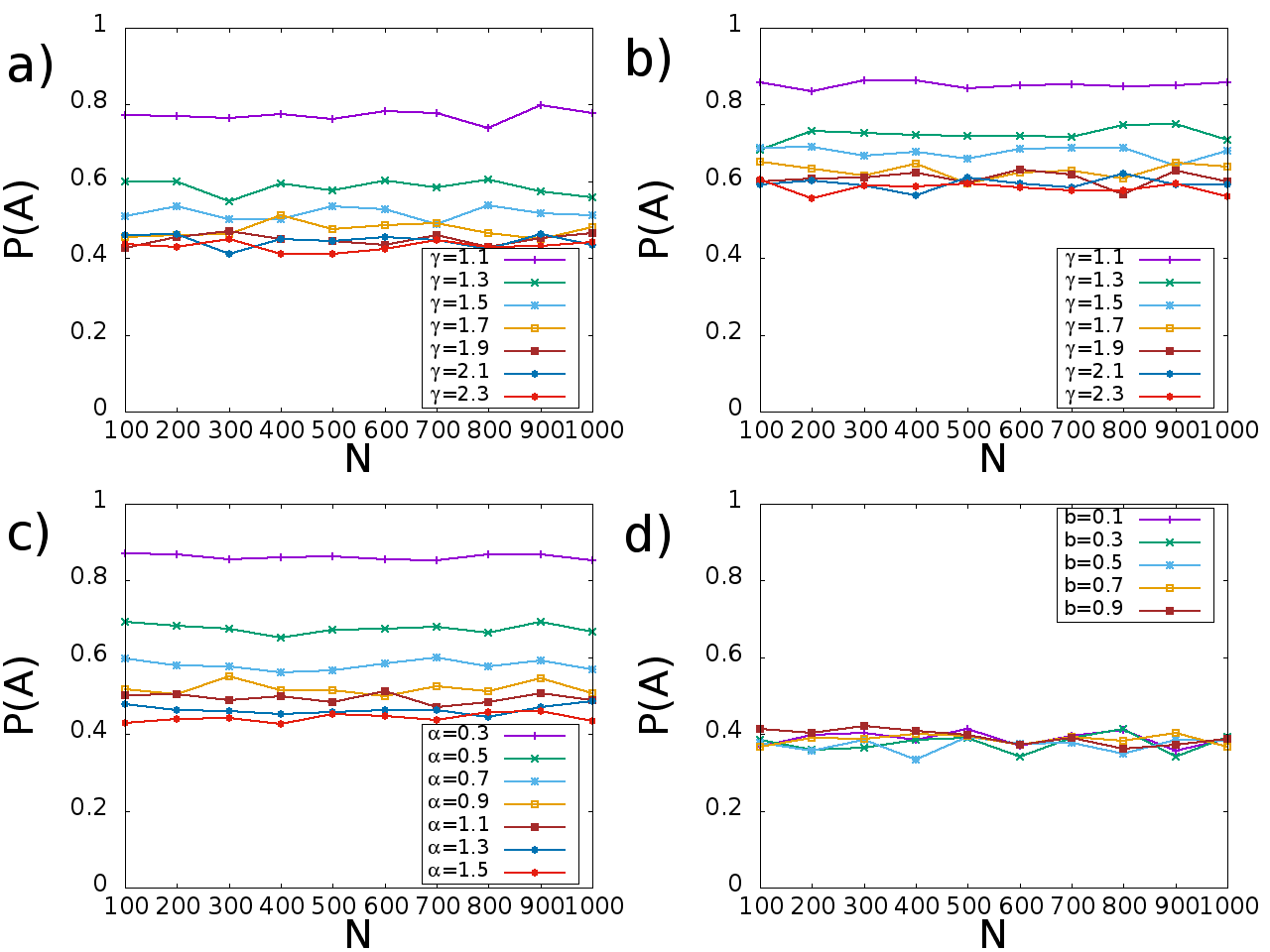}
		\caption{The fraction of runs (out of 1000 trials) vs network size that the non-Poisson ($A$-opinion) nodes won the opinion competition against the Poisson ($B$-opinion) nodes in the voter model on a complete graph. As before, the speakers' waiting-time distribution for the non-Poisson nodes is (a) power law with lower cutoff, (b) shifted power law, (c) Weibull, and (d) uniform distribution.}
		\label{voter_sims}
	\end{figure*}

	To broaden the scope of our investigations, analogous simulations were run employing the voter model \cite{Liggett1999,Castellano_review2009} on complete graphs. The voter model is very similar to the binary NG, the only difference being that it has {\em no intermediate opinion state} \cite{DallAsta2008,Vazquez2008,Zhang2014,Doyle2016}. Instead, at each time step the listener automatically accepts the speaker's opinion as its own. As before, the system is set up so that half of the nodes follow a non-Poisson update pattern and are initialized to state $A$, while the other half follow the standard Poisson pattern and are initialized to state $B$. As shown in Fig.~\ref{voter_sims}, the results are quite similar. The system remains biased towards the burstier waiting-time distribution, and the critical values of the parameters that switch the bias from the non-exponential distribution to the exponential distribution remain approximately the same. In this case, however, unlike in the binary NG, there is {\em no} symmetry breaking in the infinite system-size limit (Fig.~\ref{voter_sims}). Instead, there is a flat (system-size independent) bias towards the burstier distribution that remains the same as the system size approaches infinity. It is not clear why the voter model behaves this way in comparison to the binary NG, but it is likely the result of the lack of history-sensitivity (i.e., the lack of bistability and hysterisis) in the voter model. In the binary NG simulations, it is much harder for nodes to switch opinions. This difficulty allows for opinion shifts within the system to gain a sort of momentum as the system moves towards a consensus. In the voter model, however, the ease with which nodes change opinions means that random fluctuations are much more likely to counter all progress towards a single consensus. Thus, there is enough random noise inherent in the system that the symmetry breaking effect of heterogeneous waiting-time distributions is not as strong as in the binary NG.
	
	Note that the voter model with committed agents (on a fully-connected network) does {\em not} exhibit a tipping point in that any non-zero fraction of zealots leads to fast (exponential) relaxation to consensus \cite{Mobilia2007,Yildiz2011}. Therefore, we did not study it here.

\section{Tipping points in system of individuals with identical burstiness}

%In the main text, nearly all data was gathered on simulations with differing levels of burstiness among the competing groups, and little attention was given to the case where all nodes in the system has the same non-Poisson characteristics. For competition between two equal groups with no committed agents, the non-Poisson characteristic has no effect on the outcome. If the groups are of equal size at the start of the simulation they will each win approximately half of the simulations, and if one group is larger, it will win a larger number of the simulations, just as they would if they followed Poisson selection patterns. In the presence of committed agents, however, this changes somewhat.

\begin{figure*}
	\includegraphics[width=.75\textwidth]{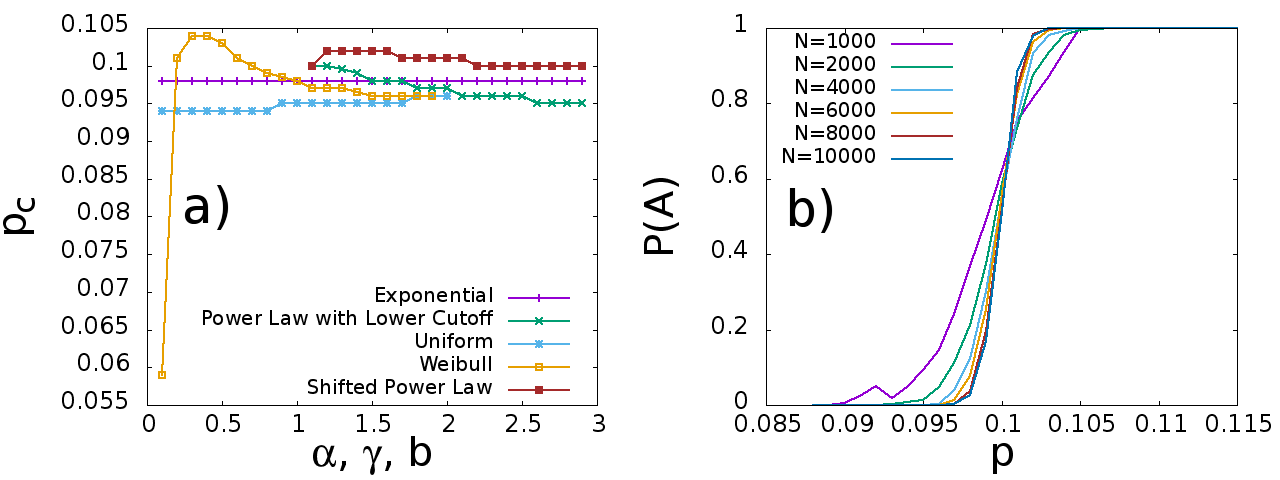}
	\caption{(a) Critical populations of committed nodes (tipping points) in the binary NG on a complete graph when each node in the network has identical waiting-time distributions and the system size is $N=1000$. (b) The Fraction of runs reaching consensus in 1000 simulations (by time $t=150$) vs the fraction of committed individuals for various system sizes. In this plot, each node has the Weibull waiting-time distribution with $\alpha=1.3$.}
	\label{pure_commit}
\end{figure*}

\begin{figure*}
	\includegraphics[width=.75\textwidth]{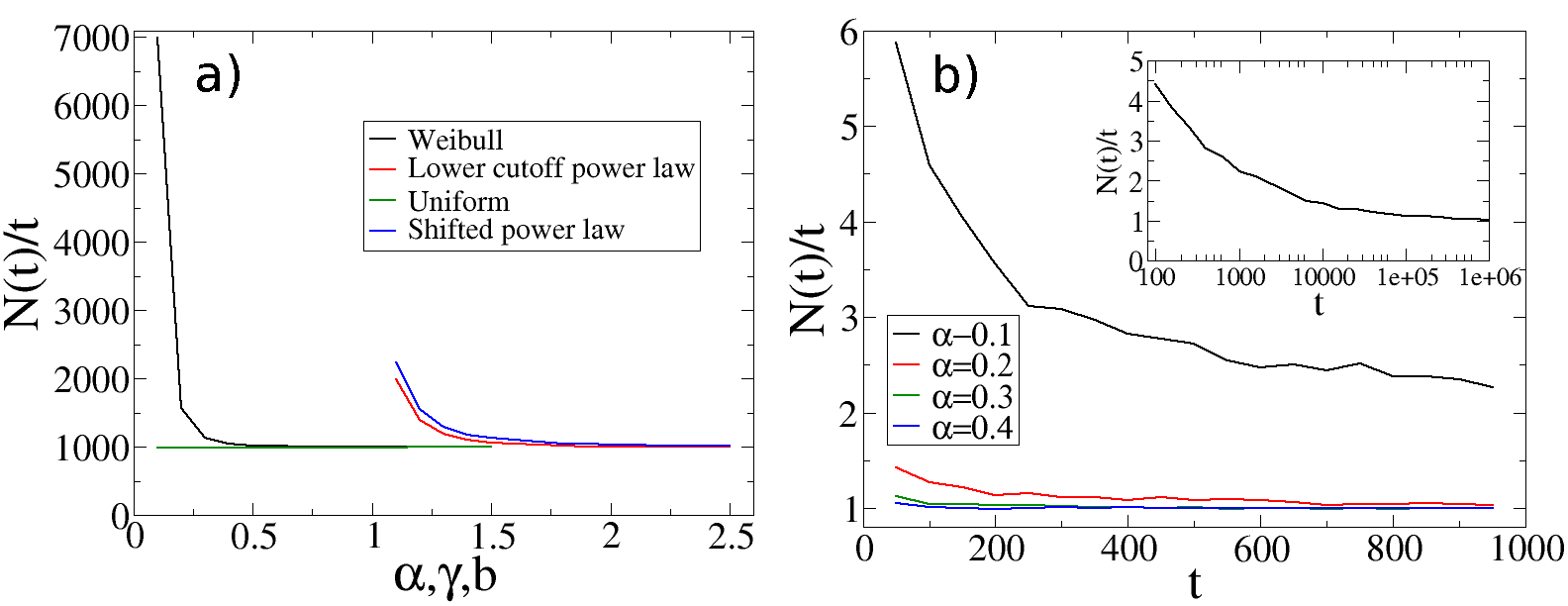}
	\caption{(a) Number of speaking events per unit system time relative to the type of waiting-time distribution used in the system for the same simulations as in Fig.~\ref{pure_commit}(a). As such, the simulations are still done on system of $N=1000$, over 1000 runs on a complete graph. (b) Average number of speakers' events per time step for a single node updating with a Weibull distributed waiting time over different time intervals. The values are for the updates of a single node averaged over 1000 simulations. The inset shows the data for $\alpha=0.1$ on extended (logarithmic) time scales}
	\label{activations_pure}
\end{figure*}

Here, we consider the case where all agents, including the committed ones, exhibit {\em identical} bursty communication characteristics [Fig.~\ref{pure_commit}]. The critical fraction of the total population (tipping points) required for fast consensus on the system exhibit some drift with regards to the waiting-time distributions used, as seen in Fig.~\ref{pure_commit}(a). In general, the critical fraction has very little dependence on the burstiness except for cases of extreme burstiness, such as a Weibull waiting-time distribution with $\alpha=0.1$. In this case, however, the effect is extreme as a result of the setup of the simulation. Each of these simulations with committed agents is set up so that there is some small fraction of individuals $p$ that is committed and in state $A$, while the rest of the network is uncommitted and in state $B$. The simulation is then run either until consensus, or until $t=150$ is reached, at which point the system is deemed as having not reached consensus. The critical population is then chosen to be the one where half of the simulations run reached consensus. This means that the system is somewhat sensitive to the value chosen for the long-time cutoff. For instance, a system left to run until $t=1000$ will return a lower value for $p_c$ because it is far more likely that somewhere in that time frame a large fluctuation will have pushed the system into consensus. The same effect can be achieved by increasing the number of speaking events per unit time $t$, yet again increasing the number of chances for a large fluctuation to occur. This is exactly what happens in this scenario, as evidenced by Fig.~\ref{activations_pure}.

Fig.~\ref{activations_pure}(a) shows that the number of speaking events per unit time in these simulations with committed agents is extremely high for the very bursty case of a Weibull waiting-time distribution with $\alpha=0.1$, but leveling out quickly for more reasonable parameter values. This is in line with what is seen in Fig.~\ref{pure_commit}(a), where the only large deviation based on burstiness is from the simulation using $\alpha=0.1$. At first glance, it is not clear why the rate should be so much higher in this case than others, considering the construction of the waiting-time distribution to have a mean of one, but Fig.~\ref{activations_pure}(b) shows that for extreme values of $\alpha$, the rate does not begin to normalize down to one until an extreme long time limit is reached. Similar results were obtained for the two power law distributions, however reaching such a ill-behaved parameter set for those distributions required values of $\gamma$ much close to the limit of one than were present in the tests in Fig.~\ref{pure_commit}. In fact, most simulations with committed agents complete in around $t\approx50$, and as mentioned above are capped at $t=150$. This is a reasonable window of study for nearly all of the distributions used, but for the most extreme cases creates an abnormally high activation rate that can skew the results.

The high rates of activations in the short times effectively explain the single large deviation in Fig.~\ref{pure_commit}(a), but also explain some of the other irregularities contained within. Upon close inspection, the distributions all either monotonically increase or decrease a very small amount, except for the Weibull and shifted power law distributions. These are the two distributions with the highest propensity for burstiness, and each has an inflection point where they change from concave to convex for values of $p_c$. This inflection point is in the same spot in Fig.~\ref{activations_pure}(a) that shows these two distribution's speaking events per unit system time normalized to the expected values. From this it can be gathered that outside of the effects of an abnormally increased speaking rate, increased burstiness works to increase the critical fraction of the population in non-Poisson update systems by a small amount (the same pattern can be seen in the systems with uniform and power law with lower cutoff distributions).

\section{Opinion Competition and the Effects of Committed Agents in the binary Naming Game on Erd\H{o}s-R\'{e}nyi Random Graphs}
	
\begin{figure*}
		\centering
		\includegraphics[width=.75\textwidth]{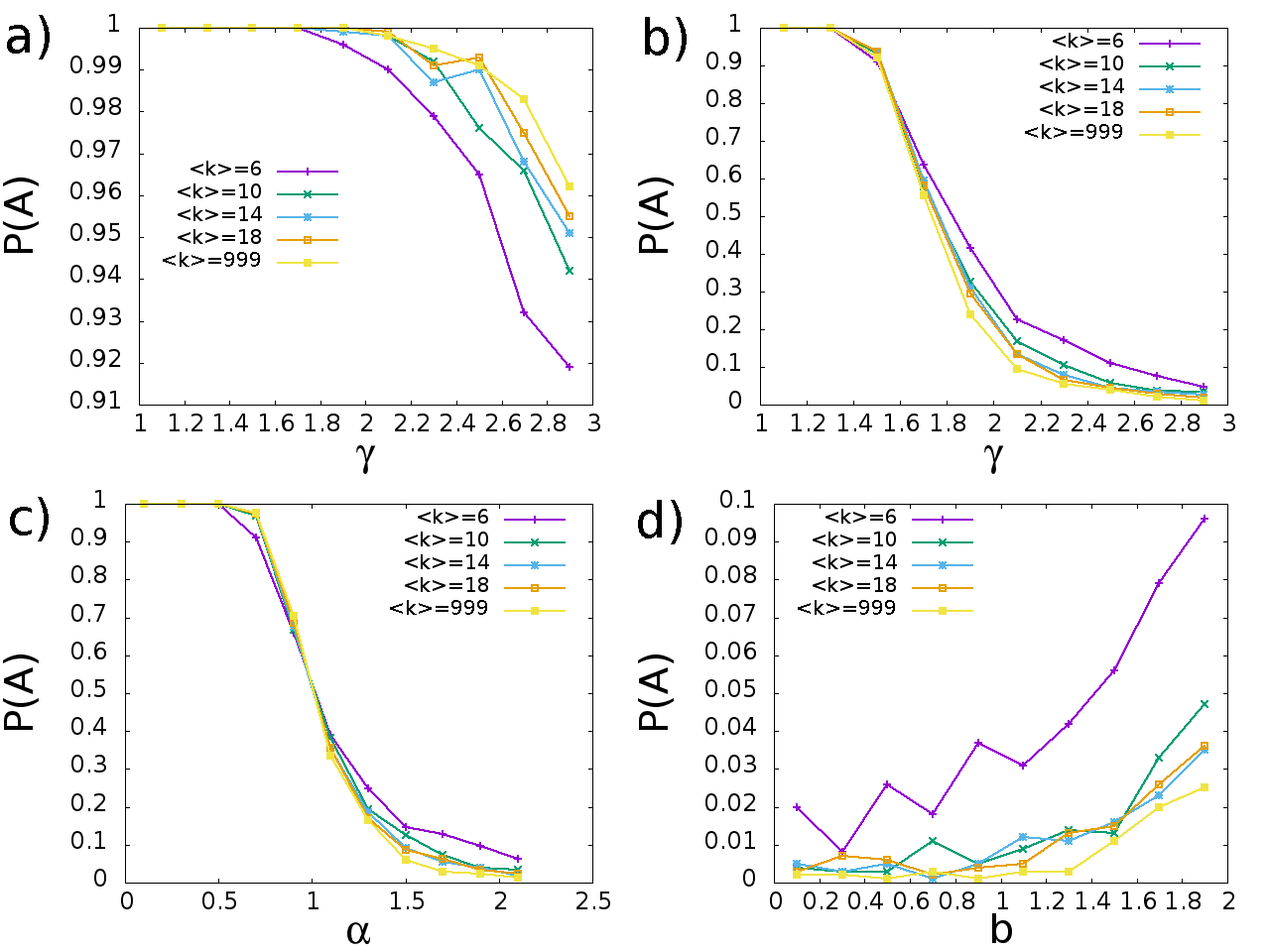}
		\caption{The fraction of runs (out of $1000$ trials) that reached consensus on opinion $A$ in ER networks with $N=1000$ nodes and various values of the average degree $\langle k\rangle$. Half of the nodes follow a non-exponential waiting-time distribution and initially have opinion $A$. The other half follow the exponential waiting-time distribution and initially have opinion $B$. The non-exponential distributions in each figure are (a) the shifted power law, (b) the power law with lower cutoff, (c) the Weibull distribution, and (d) the uniform distribution.}
		\label{er_compete}
	\end{figure*}

	In the main text all analysis is focused on the dynamics of the competition on complete graphs, but qualitatively similar effects hold in the binary NG on Erd\H{o}s-R\'{e}nyi (ER) random graphs \cite{ER1960}. In the direct competition case (seen in Fig.~\ref{er_compete}), where the simulations are initialized in the same way as in Sections \ref{ng_compete_sec}, the average degree can be seen to have minimal effect on the outcome. Having a higher average degree appears to make for a slightly more well defined transition point, but the effect is extremely small in all cases. In general, the relative burstiness at which one group can dominate the simulation is unaffected by the average degree of the network on which the system is run.
	%%%%%%%%%%%%%%%%%%%%%%%%%%%%%%%%%%%%%%%%%%%%%%%%%%%%%%%%%%%%%%%%%%%%%%%%%%%%%%%%%%%%%%%%%%%%%%%%%%%%%%%%%%%%%%%
	%%%%%%%%%%%%%%%%%%%%%%%%%%%%%%%%%%%%%%%%%%%%%%%%%%%%%%%%%%%%%%%%%%%%%%%%%%%%%%%%%%%%%%%%%%%%%%%%%%%%%%%%%%%%%%%
	
\begin{figure*}
		\centering
		\includegraphics[width=.75\textwidth]{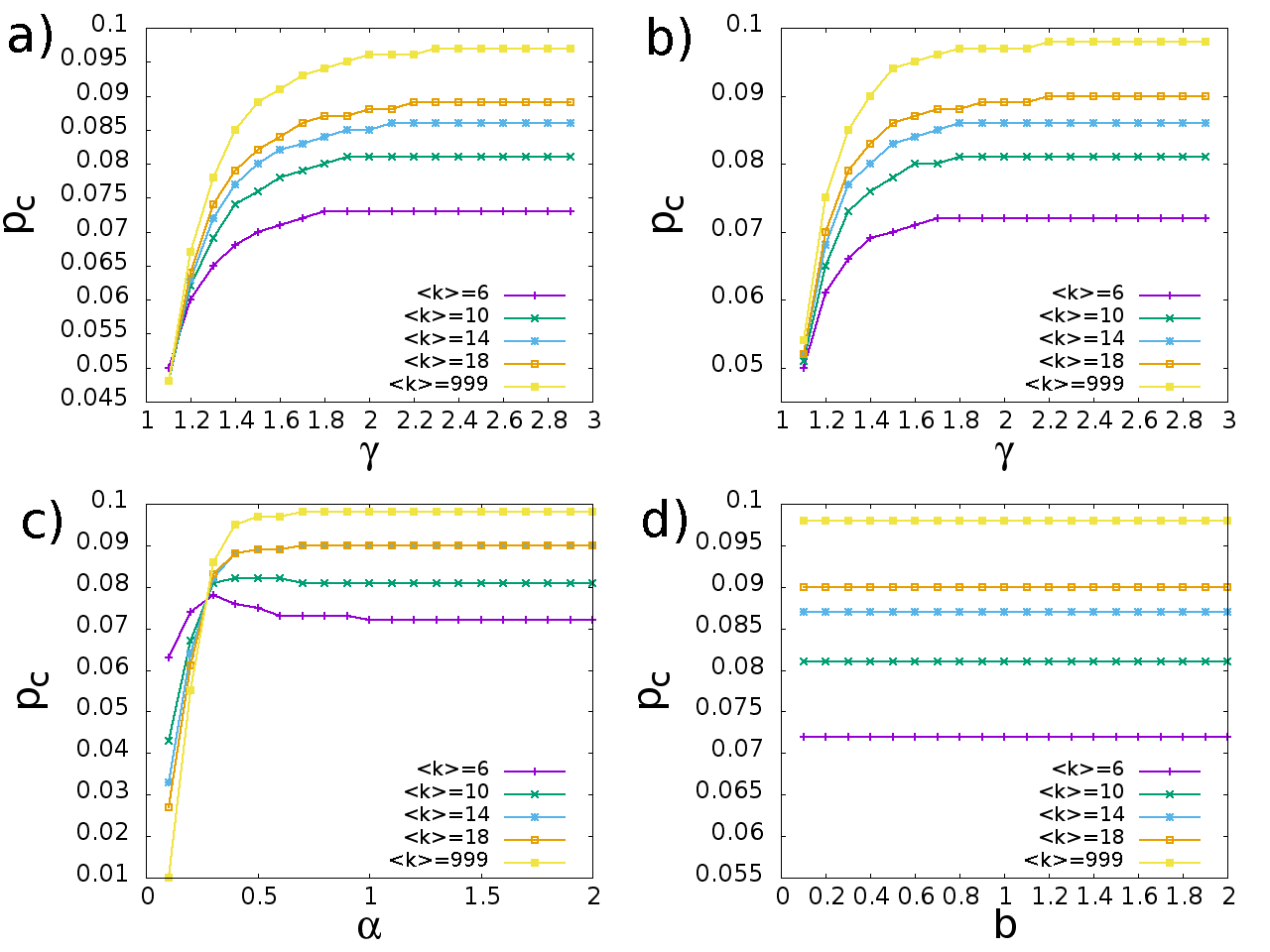}
		\caption{The critical population $p_c$ of committed nodes following a non-exponential wait time distribution that resulted in half of $1000$ trials reaching consensus on opinion $A$ in ER networks with $N=1000$ nodes and various values of the average degree $\langle k\rangle$. In each simulation a minority fraction of the population $p$ is committed to opinion $A$ and follow a non-exponential waiting time distribution. The rest of the nodes have opinion $B$ and follow the exponential distribution. The non-exponential distributions in each figure are (a) the shifted power law, (b) the power law with lower cutoff, (c) the Weibull distribution, and (d) the uniform distribution.}
		\label{er_commit}
	\end{figure*}

	Figure \ref{er_commit} shows, however, that the average degree does affect the critical population of committed agents required for fast consensus in the system. Prior works have shown that lower average degree lowers the critical population necessary for a fast consensus of the system \cite{Xie2011,Xie2012,Galehouse2014,Doyle2016}, a result that holds for systems where the nodes with non-exponential wait times are not very bursty and the system is similar to a normal naming game simulation. When high levels of burstiness are present, though, the consequent effect seems to dominate over the average degree of the network, leading to similar critical populations for many different values of $\langle k\rangle$. When taken even further into the extreme cases of burstiness (such as the Weibull distribution with $\alpha=0.1$), a lower average degree in fact raises the critical population, mitigating the effect of the extreme bursty nature of the nodes.
	%%%%%%%%%%%%%%%%%%%%%%%%%%%%%%%%%%%%%%%%%%%%%%%%%%%%%%%%%%%%%%%%%%%%%%%%%%%%%%%%%%%%%%%%%%%%%%%%%%%%%%%%%%%%%%%
	%%%%%%%%%%%%%%%%%%%%%%%%%%%%%%%%%%%%%%%%%%%%%%%%%%%%%%%%%%%%%%%%%%%%%%%%%%%%%%%%%%%%%%%%%%%%%%%%%%%%%%%%%%%%%%%%%

%
\bibliographystyle{apsrev4-1}

\newpage
\newpage


\begin{thebibliography}{34}%
\makeatletter
\providecommand \@ifxundefined [1]{%
\@ifx{#1\undefined}
}%

\providecommand \@ifnum [1]{%
\ifnum #1\expandafter \@firstoftwo
\else \expandafter \@secondoftwo
\fi
}%
\providecommand \@ifx [1]{%
\ifx #1\expandafter \@firstoftwo
\else \expandafter \@secondoftwo
\fi
}%
\providecommand \natexlab [1]{#1}%
\providecommand \enquote  [1]{``#1''}%
\providecommand \bibnamefont  [1]{#1}%
\providecommand \bibfnamefont [1]{#1}%
\providecommand \citenamefont [1]{#1}%
\providecommand \href@noop [0]{\@secondoftwo}%
\providecommand \href [0]{\begingroup \@sanitize@url \@href}%
\providecommand \@href[1]{\@@startlink{#1}\@@href}%
\providecommand \@@href[1]{\endgroup#1\@@endlink}%
\providecommand \@sanitize@url [0]{\catcode `\\12\catcode `\$12\catcode
`\&12\catcode `\#12\catcode `\^12\catcode `\_12\catcode `\%12\relax}%
\providecommand \@@startlink[1]{}%
\providecommand \@@endlink[0]{}%
\providecommand \url  [0]{\begingroup\@sanitize@url \@url }%
\providecommand \@url [1]{\endgroup\@href {#1}{\urlprefix }}%
\providecommand \urlprefix  [0]{URL }%
\providecommand \Eprint [0]{\href }%
\providecommand \doibase [0]{http://dx.doi.org/}%
\providecommand \selectlanguage [0]{\@gobble}%
\providecommand \bibinfo  [0]{\@secondoftwo}%
\providecommand \bibfield  [0]{\@secondoftwo}%
\providecommand \translation [1]{[#1]}%
\providecommand \BibitemOpen [0]{}%
\providecommand \bibitemStop [0]{}%
\providecommand \bibitemNoStop [0]{.\EOS\space}%
\providecommand \EOS [0]{\spacefactor3000\relax}%
\providecommand \BibitemShut  [1]{\csname bibitem#1\endcsname}%

\let\auto@bib@innerbib\@empty
%</preamble>


\bibitem[{\citenamefont{Castellano et~al.}(2009)\citenamefont{Castellano,
  Fortunato, and Loreto}}]{Castellano_review2009}
\bibinfo{author}{\bibfnamefont{C.}~\bibnamefont{Castellano}},
  \bibinfo{author}{\bibfnamefont{S.}~\bibnamefont{Fortunato}},
  \bibnamefont{and} \bibinfo{author}{\bibfnamefont{V.}~\bibnamefont{Loreto}},
  \bibinfo{journal}{Rev. Mod. Phys.} \textbf{\bibinfo{volume}{81}},
  \bibinfo{pages}{591} (\bibinfo{year}{2009}).

\bibitem [{\citenamefont {Galam}(2008)}]{Galam2008}%
\BibitemOpen
\bibfield  {author} {\bibinfo {author} {\bibfnamefont {S.}~\bibnamefont
{Galam}},\ }\href {\doibase 10.1142/S0129183108012297} {\bibfield  {journal}
{\bibinfo  {journal} {International Journal of Modern Physics C}\ }\textbf
{\bibinfo {volume} {19}},\ \bibinfo {pages} {409} (\bibinfo {year}
{2008})}\BibitemShut {Stop}.%

\bibitem [{\citenamefont {Castellano}\ \emph {et~al.}(2003)\citenamefont	{Castellano}, \citenamefont {Vilone},\ and\ \citenamefont {Vespignani}}]{Castellano2003}%
\BibitemOpen \bibfield  {author} {\bibinfo {author} {\bibfnamefont {C.}~\bibnamefont {Castellano}}, \bibinfo {author} {\bibfnamefont {D.}~\bibnamefont {Vilone}}, and \bibinfo {author} {\bibfnamefont {A.}~\bibnamefont {Vespignani}},}\href {\doibase 10.1209/epl/i2003-00490-0} {\bibfield  {journal} {\bibinfo{journal} { Europhysics Letters (EPL)}\ }\textbf {\bibinfo {volume} {63}},\bibinfo {pages} {153} (\bibinfo {year} {2003})}\BibitemShut {Stop}.%

\bibitem [{\citenamefont {Steels}(1995)}]{Steels1995}%
\BibitemOpen
\bibfield  {author} {\bibinfo {author} {\bibfnamefont {L.}~\bibnamefont
{Steels}},\ }\href {\doibase 10.1162/artl.1995.2.3.319} {\bibfield  {journal}
{\bibinfo  {journal} {Artificial Life}\ }\textbf {\bibinfo {volume} {2}},\
\bibinfo {pages} {319} (\bibinfo {year} {1995})}\BibitemShut {stop}.%

\bibitem [{\citenamefont {Baronchelli}\ \emph {et~al.}(2006)\citenamefont
{Baronchelli}, \citenamefont {Felici}, \citenamefont {Loreto}, \citenamefont
{Caglioti},\ and\ \citenamefont {Steels}}]{Baronchelli2006}%
\BibitemOpen
\bibfield  {author} {\bibinfo {author} {\bibfnamefont {A.}~\bibnamefont
{Baronchelli}}, \bibinfo {author} {\bibfnamefont {M.}~\bibnamefont {Felici}},
\bibinfo {author} {\bibfnamefont {V.}~\bibnamefont {Loreto}}, \bibinfo
{author} {\bibfnamefont {E.}~\bibnamefont {Caglioti}}, \ and\ \bibinfo
{author} {\bibfnamefont {L.}~\bibnamefont {Steels}},\ }\href {\doibase
10.1088/1742-5468/2006/06/P06014} {\bibfield  {journal} {\bibinfo  {journal}
{Journal of Statistical Mechanics: Theory and Experiment}\ }\textbf {\bibinfo
{volume} {2006}},\ \bibinfo {pages} {P06014} (\bibinfo {year}
{2006})}\BibitemShut {stop}.%

\bibitem [{\citenamefont {Dall'Asta}\ \emph {et~al.}(2006)\citenamefont
{Dall'Asta}, \citenamefont {Baronchelli}, \citenamefont {Barrat},\ and\
\citenamefont {Loreto}}]{DallAsta2006}%
\BibitemOpen
\bibfield  {author} {\bibinfo {author} {\bibfnamefont {L.}~\bibnamefont
{Dall'Asta}}, \bibinfo {author} {\bibfnamefont {A.}~\bibnamefont
{Baronchelli}}, \bibinfo {author} {\bibfnamefont {A.}~\bibnamefont {Barrat}},
\ and\ \bibinfo {author} {\bibfnamefont {V.}~\bibnamefont {Loreto}},\ }\href
{\doibase 10.1103/PhysRevE.74.036105} {\bibfield  {journal} {\bibinfo
{journal} {Physical Review E}\ }\textbf {\bibinfo {volume} {74}},\ \bibinfo
{pages} {036105} (\bibinfo {year} {2006})}\BibitemShut {stop}.%

\bibitem [{\citenamefont {Baronchelli}\ \emph {et~al.}(2008)\citenamefont
{Baronchelli}, \citenamefont {Loreto},\ and\ \citenamefont
{Steels}}]{Baronchelli2008}%
\BibitemOpen
\bibfield  {author} {\bibinfo {author} {\bibfnamefont {A.}~\bibnamefont
{Baronchelli}}, \bibinfo {author} {\bibfnamefont {V.}~\bibnamefont {Loreto}},
\ and\ \bibinfo {author} {\bibfnamefont {L.}~\bibnamefont {Steels}},\ }\href
{\doibase 10.1142/S0129183108012522} {\bibfield  {journal} {\bibinfo
{journal} {International Journal of Modern Physics C}\ }\textbf {\bibinfo
{volume} {19}},\ \bibinfo {pages} {785} (\bibinfo {year} {2008})}\BibitemShut
{stop}%.

\bibitem [{\citenamefont {Lu}\ \emph {et~al.}(2009)\citenamefont {Lu},
\citenamefont {Korniss},\ and\ \citenamefont {Szymanski}}]{Lu2009}%
\BibitemOpen
\bibfield  {author} {\bibinfo {author} {\bibfnamefont {Q.}~\bibnamefont
{Lu}}, \bibinfo {author} {\bibfnamefont {G.}~\bibnamefont {Korniss}}, \ and\
\bibinfo {author} {\bibfnamefont {B.~K.}\ \bibnamefont {Szymanski}},\ }\href
{\doibase 10.1007/s11403-009-0057-7} {\bibfield  {journal} {\bibinfo
{journal} {Journal of Economic Interaction and Coordination}\ }
\textbf {\bibinfo {volume} {4}},\ \bibinfo
{pages} {221} (\bibinfo {year} {2009})}
\BibitemShut {Stop}.%

\bibitem [{\citenamefont {Xie}\ \emph {et~al.}(2011)\citenamefont {Xie},
\citenamefont {Sreenivasan}, \citenamefont {Korniss}, \citenamefont {Zhang},
\citenamefont {Lim},\ and\ \citenamefont {Szymanski}}]{Xie2011}%
\BibitemOpen
\bibfield  {author} {\bibinfo {author} {\bibfnamefont {J.}~\bibnamefont
{Xie}}, \bibinfo {author} {\bibfnamefont {S.}~\bibnamefont {Sreenivasan}},
\bibinfo {author} {\bibfnamefont {G.}~\bibnamefont {Korniss}}, \bibinfo
{author} {\bibfnamefont {W.}~\bibnamefont {Zhang}}, \bibinfo {author}
{\bibfnamefont {C.}~\bibnamefont {Lim}}, \ and\ \bibinfo {author}
{\bibfnamefont {B.~K.}\ \bibnamefont {Szymanski}},\ }\href
{http://www.ncbi.nlm.nih.gov/pubmed/21867136} {\bibfield  {journal} {\bibinfo
{journal} {Physical Review. E}\ }\textbf {\bibinfo {volume} {84}},\ \bibinfo
{pages} {011130} (\bibinfo {year} {2011})}\BibitemShut {Stop}.%

\bibitem [{\citenamefont {Xie}\ \emph {et~al.}(2012)\citenamefont {Xie},
\citenamefont {Emenheiser}, \citenamefont {Kirby}, \citenamefont
{Sreenivasan}, \citenamefont {Szymanski},\ and\ \citenamefont
{Korniss}}]{Xie2012}%
\BibitemOpen
\bibfield  {author} {\bibinfo {author} {\bibfnamefont {J.}~\bibnamefont
{Xie}}, \bibinfo {author} {\bibfnamefont {J.}~\bibnamefont {Emenheiser}},
\bibinfo {author} {\bibfnamefont {M.}~\bibnamefont {Kirby}}, \bibinfo
{author} {\bibfnamefont {S.}~\bibnamefont {Sreenivasan}}, \bibinfo {author}
{\bibfnamefont {B.~K.}\ \bibnamefont {Szymanski}}, \ and\ \bibinfo {author}
{\bibfnamefont {G.}~\bibnamefont {Korniss}},\ }\href {\doibase
10.1371/journal.pone.0033215} {\bibfield  {journal} {\bibinfo  {journal}
{PloS One}\ }\textbf {\bibinfo {volume} {7}},\ \bibinfo {pages} {e33215}
(\bibinfo {year} {2012})}\BibitemShut {Stop}.%

\bibitem [{\citenamefont {Castell{\'{o}}}\ \emph {et~al.}(2009)\citenamefont
{Castell{\'{o}}}, \citenamefont {Baronchelli},\ and\ \citenamefont
{Loreto}}]{Castello2009}%
\BibitemOpen
\bibfield  {author} {\bibinfo {author} {\bibfnamefont {X.}~\bibnamefont
{Castell{\'{o}}}}, \bibinfo {author} {\bibfnamefont {A.}~\bibnamefont
{Baronchelli}}, \ and\ \bibinfo {author} {\bibfnamefont {V.}~\bibnamefont
{Loreto}},\ }\href {\doibase 10.1140/epjb/e2009-00284-2} {\bibfield
{journal} {\bibinfo  {journal} {The European Physical Journal B}\ }\textbf
{\bibinfo {volume} {71}},\ \bibinfo {pages} {557} (\bibinfo {year}
{2009})}\BibitemShut {Stop}.%

\bibitem [{\citenamefont {Karlin}\ and\ \citenamefont
	{Taylor}(1975)}]{Karlin_Taylor1975}%
\BibitemOpen
\bibfield  {author} {\bibinfo {author} {\bibfnamefont {S.}~\bibnamefont
		{Karlin}}\ and\ \bibinfo {author} {\bibfnamefont {H.}~\bibnamefont
		{Taylor}},\ }\href@noop {} {\emph {\bibinfo {title} {A First Course in
			Stochastic Processes, Second Edition}}}\ (\bibinfo  {publisher} {Academic
	Press},\ \bibinfo {address} {New York},\ \bibinfo {year} {1975})\BibitemShut
{Stop}.%

\bibitem [{\citenamefont {Candia}\ \emph {et~al.}(2008)\citenamefont {Candia},
\citenamefont {Gonz{\'{a}}lez}, \citenamefont {Wang}, \citenamefont
{Schoenharl}, \citenamefont {Madey},\ and\ \citenamefont
{Barab{\'{a}}si}}]{Candia2008}%
\BibitemOpen
\bibfield  {author} {\bibinfo {author} {\bibfnamefont {J.}~\bibnamefont
{Candia}}, \bibinfo {author} {\bibfnamefont {M.~C.}\ \bibnamefont
{Gonz{\'{a}}lez}}, \bibinfo {author} {\bibfnamefont {P.}~\bibnamefont
{Wang}}, \bibinfo {author} {\bibfnamefont {T.}~\bibnamefont {Schoenharl}},
\bibinfo {author} {\bibfnamefont {G.}~\bibnamefont {Madey}}, \ and\ \bibinfo
{author} {\bibfnamefont {A.-L.}\ \bibnamefont {Barab{\'{a}}si}},\ }\href
{\doibase 10.1088/1751-8113/41/22/224015} {\bibfield  {journal} {\bibinfo
{journal} {Journal of Physics A: Mathematical and Theoretical}\ }\textbf
{\bibinfo {volume} {41}},\ \bibinfo {pages} {224015} (\bibinfo {year}
{2008})}\BibitemShut {Stop}.%

\bibitem [{\citenamefont {Barab{\'{a}}si}(2005)}]{Barabasi2005}%
\BibitemOpen
\bibfield  {author} {\bibinfo {author} {\bibfnamefont {A.-L.}\ \bibnamefont
{Barab{\'{a}}si}},\ }\href {\doibase 10.1038/nature03459} {\bibfield
{journal} {\bibinfo  {journal} {Nature}\ }\textbf {\bibinfo {volume} {435}},\
\bibinfo {pages} {207} (\bibinfo {year} {2005})}\BibitemShut {NoStop}%
\bibitem [{\citenamefont {V{\'{a}}zquez}\ \emph {et~al.}(2006)\citenamefont
{V{\'{a}}zquez}, \citenamefont {Oliveira}, \citenamefont {Dezs{\"{o}}},
\citenamefont {Goh}, \citenamefont {Kondor},\ and\ \citenamefont
{Barab{\'{a}}si}}]{Vazquez2006}%
\BibitemOpen
\bibfield  {author} {\bibinfo {author} {\bibfnamefont {A.}~\bibnamefont
{V{\'{a}}zquez}}, \bibinfo {author} {\bibfnamefont {J.~G.}\ \bibnamefont
{Oliveira}}, \bibinfo {author} {\bibfnamefont {Z.}~\bibnamefont
{Dezs{\"{o}}}}, \bibinfo {author} {\bibfnamefont {K.-I.}\ \bibnamefont
{Goh}}, \bibinfo {author} {\bibfnamefont {I.}~\bibnamefont {Kondor}}, \ and\
\bibinfo {author} {\bibfnamefont {A.-L.}\ \bibnamefont {Barab{\'{a}}si}},\
}\href {\doibase 10.1103/PhysRevE.73.036127} {\bibfield  {journal} {\bibinfo
{journal} {Physical review. E, Statistical, nonlinear, and soft matter physics}\ }\textbf {\bibinfo {volume} {73}},\ \bibinfo {pages} {036127}
(\bibinfo {year} {2006})}\BibitemShut {Stop}.%

\bibitem [{\citenamefont {Goh}\ and\ \citenamefont
{Barab{\'{a}}si}(2008)}]{Goh2008}%
\BibitemOpen
\bibfield  {author} {\bibinfo {author} {\bibfnamefont {K.}~\bibnamefont
{Goh}}\ and\ \bibinfo {author} {\bibfnamefont {A.}~\bibnamefont
{Barab{\'{a}}si}},\ }\href {\doibase 10.1209/0295-5075/81/48002} {\bibfield
{journal} {\bibinfo  {journal} {Europhys Lett}\ }\textbf {\bibinfo {volume}
{81}},\ \bibinfo {pages} {48002} (\bibinfo {year} {2008})}\BibitemShut
{Stop}.%

\bibitem [{\citenamefont {Karsai}\ \emph {et~al.}(2012)\citenamefont {Karsai},
\citenamefont {Kaski}, \citenamefont {Barab{\'{a}}si},\ and\ \citenamefont
{Kert{\'{e}}sz}}]{Karsai2012}%
\BibitemOpen
\bibfield  {author} {\bibinfo {author} {\bibfnamefont {M.}~\bibnamefont
{Karsai}}, \bibinfo {author} {\bibfnamefont {K.}~\bibnamefont {Kaski}},
\bibinfo {author} {\bibfnamefont {A.-L.}\ \bibnamefont {Barab{\'{a}}si}}, \
and\ \bibinfo {author} {\bibfnamefont {J.}~\bibnamefont {Kert{\'{e}}sz}},\
}\href {\doibase 10.1038/srep00397} {\bibfield  {journal} {\bibinfo
{journal} {Scientific Reports}\ }\textbf {\bibinfo {volume} {2}},\ \bibinfo
{pages} {397} (\bibinfo {year} {2012})}\BibitemShut {Stop}.%

\bibitem [{\citenamefont {Artime}\ \emph {et~al.}(2016)\citenamefont {Artime},
	\citenamefont {Ramasco},\ and\ \citenamefont {Miguel}}]{Artime2016}%
\BibitemOpen
\bibfield  {author} {\bibinfo {author} {\bibfnamefont {O.}~\bibnamefont
		{Artime}}, \bibinfo {author} {\bibfnamefont {J.~J.}\ \bibnamefont {Ramasco}},
	\ and\ \bibinfo {author} {\bibfnamefont {M.~S.}\ \bibnamefont {Miguel}},\
}\href {http://arxiv.org/abs/1604.04155} {\  (\bibinfo {year} {2016})},\
\Eprint {http://arxiv.org/abs/1604.04155} {arXiv:1604.04155} \BibitemShut
{Stop}.%

\bibitem [{\citenamefont {Doyle}\ \emph {et~al.}(2016)\citenamefont {Doyle},
	\citenamefont {Sreenivasan}, \citenamefont {Szymanski},\ and\ \citenamefont
	{Korniss}}]{Doyle2016}%
\BibitemOpen
\bibfield  {author} {\bibinfo {author} {\bibfnamefont {C.}~\bibnamefont
		{Doyle}}, \bibinfo {author} {\bibfnamefont {S.}~\bibnamefont {Sreenivasan}},
	\bibinfo {author} {\bibfnamefont {B.}~\bibnamefont {Szymanski}}, \ and\
	\bibinfo {author} {\bibfnamefont {G.}~\bibnamefont {Korniss}},\ }\href
{\doibase 10.1016/j.physa.2015.09.081} {\bibfield  {journal} {\bibinfo
		{journal} {Physica A: Statistical Mechanics and its Applications}\ }\textbf
	{\bibinfo {volume} {443}},\ \bibinfo {pages} {316} (\bibinfo {year}
	{2016})}\BibitemShut {Stop}.%

\bibitem [{\citenamefont {Zhang}\ \emph {et~al.}(2011)\citenamefont {Zhang},
	\citenamefont {Lim}, \citenamefont {Sreenivasan}, \citenamefont {Xie},
	\citenamefont {Szymanski},\ and\ \citenamefont {Korniss}}]{Zhang2011}%
\BibitemOpen
\bibfield  {author} {\bibinfo {author} {\bibfnamefont {W.}~\bibnamefont
		{Zhang}}, \bibinfo {author} {\bibfnamefont {C.}~\bibnamefont {Lim}}, \bibinfo
	{author} {\bibfnamefont {S.}~\bibnamefont {Sreenivasan}}, \bibinfo {author}
	{\bibfnamefont {J.}~\bibnamefont {Xie}}, \bibinfo {author} {\bibfnamefont
		{B.~K.}\ \bibnamefont {Szymanski}}, \ and\ \bibinfo {author} {\bibfnamefont
		{G.}~\bibnamefont {Korniss}},\ }\href {\doibase 10.1063/1.3598450} {\bibfield
	{journal} {\bibinfo  {journal} {Chaos: An Interdisciplinary Journal of
			Nonlinear Science}\ }\textbf {\bibinfo {volume} {21}},\ \bibinfo {pages}
	{025115} (\bibinfo {year} {2011})}\BibitemShut {Stop}.%


\bibitem [{\citenamefont {Galehouse}\ \emph {et~al.}(2014)\citenamefont
	{Galehouse}, \citenamefont {Nguyen}, \citenamefont {Sreenivasan},
	\citenamefont {Lizardo}, \citenamefont {Korniss},\ and\ \citenamefont
	{Szymanski}}]{Galehouse2014}%
\BibitemOpen
\bibfield  {author} {\bibinfo {author} {\bibfnamefont {D.}~\bibnamefont
		{Galehouse}}, \bibinfo {author} {\bibfnamefont {T.}~\bibnamefont {Nguyen}},
	\bibinfo {author} {\bibfnamefont {S.}~\bibnamefont {Sreenivasan}}, \bibinfo
	{author} {\bibfnamefont {O.}~\bibnamefont {Lizardo}}, \bibinfo {author}
	{\bibfnamefont {G.}~\bibnamefont {Korniss}}, \ and\ \bibinfo {author}
	{\bibfnamefont {B.}~\bibnamefont {Szymanski}},\ }\href@noop {} {\bibfield
	{journal} {\bibinfo  {journal} {Proc. 5th International Conference on Applied
			Human Factors and Ergonomics AFHE, Krakaw, Poland, July 29-23}\ ,\ \bibinfo
		{pages} {2318}} (\bibinfo {year} {2014})}\BibitemShut {Stop}.%

\bibitem [{\citenamefont {Liu}\ \emph {et~al.}(2012)\citenamefont {Liu},
	\citenamefont {Wu},\ and\ \citenamefont {Zhang}}]{Liu2012}%
\BibitemOpen
\bibfield  {author} {\bibinfo {author} {\bibfnamefont {X.-T.}\ \bibnamefont
		{Liu}}, \bibinfo {author} {\bibfnamefont {Z.-X.}\ \bibnamefont {Wu}}, \ and\
	\bibinfo {author} {\bibfnamefont {L.}~\bibnamefont {Zhang}},\ }\href
{\doibase 10.1103/PhysRevE.86.051132} {\bibfield  {journal} {\bibinfo
		{journal} {Physical Review E}\ }\textbf {\bibinfo {volume} {86}},\ \bibinfo
	{pages} {051132} (\bibinfo {year} {2012})}\BibitemShut {Stop}.%

\bibitem [{\citenamefont {Thompson}\ \emph {et~al.}(2014)\citenamefont
	{Thompson}, \citenamefont {Szymanski},\ and\ \citenamefont
	{Lim}}]{Thompson2014}%
\BibitemOpen
\bibfield  {author} {\bibinfo {author} {\bibfnamefont {A.~M.}\ \bibnamefont
		{Thompson}}, \bibinfo {author} {\bibfnamefont {B.~K.}\ \bibnamefont
		{Szymanski}}, \ and\ \bibinfo {author} {\bibfnamefont {C.~C.}\ \bibnamefont
		{Lim}},\ }\href {\doibase 10.1103/PhysRevE.90.042809} {\bibfield  {journal}
	{\bibinfo  {journal} {Physical Review E}\ }\textbf {\bibinfo {volume} {90}},\
	\bibinfo {pages} {042809} (\bibinfo {year} {2014})}\BibitemShut {Stop}.%

\bibitem [{\citenamefont {Zhang}\ \emph {et~al.}(2012)\citenamefont {Zhang},
	\citenamefont {Lim},\ and\ \citenamefont {Szymanski}}]{Zhang2012}%
\BibitemOpen
\bibfield  {author} {\bibinfo {author} {\bibfnamefont {W.}~\bibnamefont
		{Zhang}}, \bibinfo {author} {\bibfnamefont {C.}~\bibnamefont {Lim}}, \ and\
	\bibinfo {author} {\bibfnamefont {B.~K.}\ \bibnamefont {Szymanski}},\ }\href
{\doibase 10.1103/PhysRevE.86.061134} {\bibfield  {journal} {\bibinfo
		{journal} {Physical Review E}\ }\textbf {\bibinfo {volume} {86}}, \bibinfo
	{pages} {061134} (\bibinfo {year} {2012})}\BibitemShut{Stop}.%

\bibitem [{\citenamefont {Mobilia}(2003)}]{Mobilia2003}%
\BibitemOpen
\bibfield  {author} {\bibinfo {author} {\bibfnamefont {M.}~\bibnamefont {Mobilia}},\ }\href {\doibase 10.1103/PhysRevLett.91.028701} {\bibfield {journal} {\bibinfo  {journal} {Physical Review Letters}\ }\textbf {\bibinfo {volume} {91}},\ \bibinfo {pages} {028701} (\bibinfo {year} {2003})}\BibitemShut {Stop}.%

\bibitem [{\citenamefont {Mobilia}\ \emph {et~al.}(2007)\citenamefont {Mobilia}, \citenamefont {Petersen}, \citenamefont {Redner},}]{Mobilia2007}%
\BibitemOpen
\bibfield  {author} {\bibinfo {author} {\bibfnamefont {M.}~\bibnamefont {Mobilia}}, \bibinfo {author} {\bibfnamefont {A.}~\bibnamefont {Petersen}}, \bibinfo {author} {\bibfnamefont {S.}~\bibnamefont {Redner}},\ }\href {\doibase 10.1088/1742-5468/2007/08/P08029} {\bibfield  {journal} {\bibinfo  {journal} {Journal of Statistical Mechanics: Theory and Experiment}\ }\textbf {\bibinfo {volume} {2007}}, \bibinfo {pages} {P08029} (\bibinfo {year} {2007})}\BibitemShut {Stop}.%

\bibitem [{\citenamefont {Galam}\ and\ \citenamefont {Jacobs}(2007)}]{Galam2007}%
\BibitemOpen
\bibfield  {author} {\bibinfo {author} {\bibfnamefont {S.}~\bibnamefont
		{Galam}}\ and\ \bibinfo {author} {\bibfnamefont {F.}~\bibnamefont {Jacobs}},\
}\href {\doibase 10.1016/j.physa.2007.03.034} {\bibfield  {journal} {\bibinfo
	{journal} {Physica A: Statistical Mechanics and its Applications}\ }\textbf
{\bibinfo {volume} {381}},\ \bibinfo {pages} {366} (\bibinfo {year}
{2007})}\BibitemShut {Stop}.%

\bibitem [{\citenamefont {Yildiz}\ \emph {et~al.}(2011)\citenamefont {Yildiz},
	\citenamefont {Acemoglu}, \citenamefont {Ozdaglar}, \citenamefont {Saberi},\
	and\ \citenamefont {Scaglione}}]{Yildiz2011}%
\BibitemOpen
\bibfield  {author} {\bibinfo {author} {\bibfnamefont {E.}~\bibnamefont
		{Yildiz}}, \bibinfo {author} {\bibfnamefont {D.}~\bibnamefont {Acemoglu}},
	\bibinfo {author} {\bibfnamefont {A.~E.}\ \bibnamefont {Ozdaglar}}, \bibinfo
	{author} {\bibfnamefont {A.}~\bibnamefont {Saberi}}, \ and\ \bibinfo {author}
	{\bibfnamefont {A.}~\bibnamefont {Scaglione}},\ }\href {\doibase
	10.2139/ssrn.1744113} {\bibfield  {journal} {\bibinfo  {journal} {SSRN
			Electronic Journal}\ } (\bibinfo {year} {2011}),\
	10.2139/ssrn.1744113}\BibitemShut {Stop}.%

\bibitem [{\citenamefont {Marvel}\ \emph {et~al.}(2012)\citenamefont {Marvel},
	\citenamefont {Hong}, \citenamefont {Papush},\ and\ \citenamefont
	{Strogatz}}]{Marvel2012}%
\BibitemOpen
\bibfield  {author} {\bibinfo {author} {\bibfnamefont {S.~A.}\ \bibnamefont
		{Marvel}}, \bibinfo {author} {\bibfnamefont {H.}~\bibnamefont {Hong}},
	\bibinfo {author} {\bibfnamefont {A.}~\bibnamefont {Papush}}, \ and\ \bibinfo
	{author} {\bibfnamefont {S.~H.}\ \bibnamefont {Strogatz}},\ }\href {\doibase
	10.1103/PhysRevLett.109.118702} {\bibfield  {journal} {\bibinfo  {journal}
		{Physical Review Letters}\ }\textbf {\bibinfo {volume} {109}},\ \bibinfo
	{pages} {118702} (\bibinfo {year} {2012})}\BibitemShut {Stop}.%

\bibitem{Verma2014}
G. Verma, A. Swami, and K. Chan,
%"The impact of competing zealots on opinion dynamics",
Physica A {\bf 395}, 310 (2014).

\bibitem{Waagen2015}
A. Waagen, G. Verma, K. Chan, A. Swami, and R. D'Souza,
%"Effect of zealotry in high-dimensional opinion dynamics models",
Phys. Rev. E {\bf 91}, 022811 (2015).

\bibitem [{\citenamefont {Vazquez}\ \emph {et~al.}(2007)\citenamefont
{Vazquez}, \citenamefont {R{\'{a}}cz}, \citenamefont {Luk{\'{a}}cs},\ and\
\citenamefont {Barab{\'{a}}si}}]{Vazquez2007}%
\BibitemOpen
\bibfield  {author} {\bibinfo {author} {\bibfnamefont {A.}~\bibnamefont
{Vazquez}}, \bibinfo {author} {\bibfnamefont {B.}~\bibnamefont {R{\'{a}}cz}},
\bibinfo {author} {\bibfnamefont {A.}~\bibnamefont {Luk{\'{a}}cs}}, \ and\
\bibinfo {author} {\bibfnamefont {A.-L.}\ \bibnamefont {Barab{\'{a}}si}},\
}\href {\doibase 10.1103/PhysRevLett.98.158702} {\bibfield  {journal}
{\bibinfo  {journal} {Physical Review Letters}\ }\textbf {\bibinfo {volume}
{98}},\ \bibinfo {pages} {158702} (\bibinfo {year} {2007})}\BibitemShut
{Stop}.%

\bibitem [{\citenamefont {Karsai}\ \emph {et~al.}(2011)\citenamefont {Karsai},
\citenamefont {Kivel{\"{a}}}, \citenamefont {Pan}, \citenamefont {Kaski},
\citenamefont {Kert{\'{e}}sz}, \citenamefont {Barab{\'{a}}si},\ and\
\citenamefont {Saram{\"{a}}ki}}]{Karsai2011}%
\BibitemOpen
\bibfield  {author} {\bibinfo {author} {\bibfnamefont {M.}~\bibnamefont
{Karsai}}, \bibinfo {author} {\bibfnamefont {M.}~\bibnamefont
{Kivel{\"{a}}}}, \bibinfo {author} {\bibfnamefont {R.~K.}\ \bibnamefont
{Pan}}, \bibinfo {author} {\bibfnamefont {K.}~\bibnamefont {Kaski}}, \bibinfo
{author} {\bibfnamefont {J.}~\bibnamefont {Kert{\'{e}}sz}}, \bibinfo {author}
{\bibfnamefont {A.-L.}\ \bibnamefont {Barab{\'{a}}si}}, \ and\ \bibinfo
{author} {\bibfnamefont {J.}~\bibnamefont {Saram{\"{a}}ki}},\ }\href
{\doibase 10.1103/PhysRevE.83.025102} {\bibfield  {journal} {\bibinfo
{journal} {Physical Review E}\ }\textbf {\bibinfo {volume} {83}},\ \bibinfo
{pages} {025102} (\bibinfo {year} {2011})}\BibitemShut {Stop}.%

\bibitem [{\citenamefont {Iribarren}\ and\ \citenamefont
{Moro}(2009)}]{Iribarren2009}%
\BibitemOpen
\bibfield  {author} {\bibinfo {author} {\bibfnamefont {J.}~\bibnamefont
{Iribarren}}\ and\ \bibinfo {author} {\bibfnamefont {E.}~\bibnamefont
{Moro}},\ }\href {\doibase 10.1103/PhysRevLett.103.038702} {\bibfield
{journal} {\bibinfo  {journal} {Physical Review Letters}\ }\textbf {\bibinfo
{volume} {103}},\ \bibinfo {pages} {038702} (\bibinfo {year}
{2009})}\BibitemShut {Stop}.%

\bibitem [{\citenamefont {Lambiotte}\ \emph {et~al.}(2013)\citenamefont
{Lambiotte}, \citenamefont {Tabourier},\ and\ \citenamefont
{Delvenne}}]{Lambiotte2013}%
\BibitemOpen
\bibfield  {author} {\bibinfo {author} {\bibfnamefont {R.}~\bibnamefont
{Lambiotte}}, \bibinfo {author} {\bibfnamefont {L.}~\bibnamefont
{Tabourier}}, \ and\ \bibinfo {author} {\bibfnamefont {J.-C.}\ \bibnamefont
{Delvenne}},\ }\href {\doibase 10.1140/epjb/e2013-40456-9} {\bibfield
{journal} {\bibinfo  {journal} {The European Physical Journal B}\ }\textbf
{\bibinfo {volume} {86}},\ \bibinfo {pages} {320} (\bibinfo {year}
{2013})}\BibitemShut {Stop}.%

\bibitem [{\citenamefont {Takaguchi}\ \emph {et~al.}(2013)\citenamefont
{Takaguchi}, \citenamefont {Masuda},\ and\ \citenamefont
{Holme}}]{Takaguchi2013}%
\BibitemOpen
\bibfield  {author} {\bibinfo {author} {\bibfnamefont {T.}~\bibnamefont
{Takaguchi}}, \bibinfo {author} {\bibfnamefont {N.}~\bibnamefont {Masuda}}, \
and\ \bibinfo {author} {\bibfnamefont {P.}~\bibnamefont {Holme}},\ }\href
{\doibase 10.1371/journal.pone.0068629} {\bibfield  {journal} {\bibinfo
{journal} {PloS one}\ }\textbf {\bibinfo {volume} {8}},\ \bibinfo {pages}
{e68629} (\bibinfo {year} {2013})}\BibitemShut {Stop}.%

\bibitem [{\citenamefont {{Van Mieghem}}\ and\ \citenamefont {van~de
Bovenkamp}(2013)}]{VanMieghem2013}%
\BibitemOpen
\bibfield  {author} {\bibinfo {author} {\bibfnamefont {P.}~\bibnamefont {{Van
Mieghem}}}\ and\ \bibinfo {author} {\bibfnamefont {R.}~\bibnamefont {van~de
Bovenkamp}},\ }\href {\doibase 10.1103/PhysRevLett.110.108701} {\bibfield
{journal} {\bibinfo  {journal} {Physical Review Letters}\ }\textbf {\bibinfo
{volume} {110}},\ \bibinfo {pages} {108701} (\bibinfo {year}
{2013})}\BibitemShut {Stop}.%

\bibitem [{\citenamefont {Mistry}\ \emph {et~al.}(2015)\citenamefont {Mistry},
	\citenamefont {Zhang}, \citenamefont {Perra},\ and\ \citenamefont
	{Baronchelli}}]{Mistry2015}%
\BibitemOpen
\bibfield  {author} {\bibinfo {author} {\bibfnamefont {D.}~\bibnamefont
		{Mistry}}, \bibinfo {author} {\bibfnamefont {Q.}~\bibnamefont {Zhang}},
	\bibinfo {author} {\bibfnamefont {N.}~\bibnamefont {Perra}}, \ and\ \bibinfo
	{author} {\bibfnamefont {A.}~\bibnamefont {Baronchelli}},\ }\href {\doibase
	10.1103/PhysRevE.92.042805} {\bibfield  {journal} {\bibinfo  {journal}
		{Physical Review E}\ }\textbf {\bibinfo {volume} {92}},\ \bibinfo {pages}
	{042805} (\bibinfo {year} {2015})}\BibitemShut {Stop}.%


\bibitem{ER1960}
Erd\H{o}s-R\'{e}nyi,
%``On the evolution of random graphs",
Publ. Math. Inst. Hung. Acad. Sci. {\bf 5}, 17 (1960).



\bibitem [{\citenamefont {Johnson}\ \emph {et~al.}(1994)\citenamefont
{Johnson}, \citenamefont {Kotz},\ and\ \citenamefont
{Balakrishnan}}]{Johnson1994}%
\BibitemOpen
\bibfield  {author} {\bibinfo {author} {\bibfnamefont {N.~L.}\ \bibnamefont
{Johnson}}, \bibinfo {author} {\bibfnamefont {S.}~\bibnamefont {Kotz}}, \
and\ \bibinfo {author} {\bibfnamefont {N.}~\bibnamefont {Balakrishnan}},\
}\href@noop {} {\emph {\bibinfo {title} {Continuous Univariate Distributions,
Volume 1, 2nd Edition}}}\ (\bibinfo  {publisher} {Wiley-Interscience},\
\bibinfo {year} {1994})\BibitemShut {NoStop}%
\bibitem [{\citenamefont {White}(1964)}]{White1964}%
\BibitemOpen
\bibfield  {author} {\bibinfo {author} {\bibfnamefont {J.~S.}\ \bibnamefont
{White}},\ }in\ \href {\doibase 10.4271/640624} {\emph {\bibinfo {booktitle}
{SAE Technical Paper}}}\ (\bibinfo  {publisher} {SAE International},\
\bibinfo {year} {1964})\BibitemShut {Stop}.%


\bibitem{Liggett1999}
T.M. Liggett, Stochastic Interacting Systems: Contact, Voter and Exclusion Processes, Springer-Verlag, 1999.

\bibitem{DallAsta2008}
L. Dall'Asta and T. Galla,
%Algebraic coarsening in voter models with intermediate states,
J. Phys. A: Mathematical and Theoretical 41 (2008) 435003.

\bibitem{Vazquez2008}
F. Vazquez and C. Lopez,
%Systems with two symmetric absorbing states: relating the microscopic dynamics with the macroscopic behavior,
Phys. Rev. E 78 (2008) 061127.

\bibitem{Zhang2014}
W. Zhang, C. Lim, G. Korniss and B.K. Szymanski,
Sci. Rep. 4 (2014) 5568.

\bibitem [{{\relax DLMF}()}]{DLMF}%
\BibitemOpen
{\relax DLMF},\ \href {http://dlmf.nist.gov/} {\enquote {\bibinfo {title}
		{{\it NIST Digital Library of Mathematical Functions}},}\ }\bibinfo
{howpublished} {http://dlmf.nist.gov/, Release 1.0.14 of 2016-12-21},\
\bibinfo {note} {f.~W.~J. Olver, A.~B. {Olde Daalhuis}, D.~W. Lozier, B.~I.
	Schneider, R.~F. Boisvert, C.~W. Clark, B.~R. Miller and B.~V. Saunders,
	eds.}\BibitemShut {Stop}.%

\end{thebibliography}

\end{document}